\documentclass[12pt]{article}
\usepackage{color,amssymb,epsfig,verbatim,amsmath}
\usepackage[ruled,vlined]{algorithm2e}
\usepackage[title]{appendix}

\def\boxit#1{\vbox{\hrule\hbox{\vrule\kern6pt
          \vbox{\kern6pt#1\kern6pt}\kern6pt\vrule}\hrule}}
\def\refhg{\hangindent=20pt\hangafter=1}
\def\refmark{\par\vskip 2mm\noindent\refhg}

\def\refhg{\hangindent=20pt\hangafter=1}
\def\refmark{\par\vskip 2mm\noindent\refhg}

\def\bse{\begin{eqnarray*}}
\def\ese{\end{eqnarray*}}
\def\be{\begin{eqnarray}}
\def\ee{\end{eqnarray}}
\def\bq{\begin{equation}}
\def\eq{\end{equation}}
\def\bse{\begin{eqnarray*}}
\def\ese{\end{eqnarray*}}

\usepackage{pgf,tikz}
\usetikzlibrary{arrows}
\usepackage{hyperref}
\hypersetup{
    colorlinks=true,
    linkcolor=blue,
    filecolor=magenta,      
    urlcolor=blue,
}
\usepackage{lscape}
\usepackage{pdflscape}
\usepackage{afterpage}
\usepackage{longtable,booktabs}
\usepackage{multirow}
\usepackage{threeparttable}
\usepackage[ruled,vlined]{algorithm2e}

\newtheorem{remark}{Remark}[section]
\newtheorem{theorem}{Theorem}[section]

\makeatletter
\newcommand*\bigcdot{\mathpalette\bigcdot@{.5}}
\newcommand*\bigcdot@[2]{\mathbin{\vcenter{\hbox{\scalebox{#2}{$\m@th#1\bullet$}}}}}
\makeatother
\begin{document}
\thispagestyle{empty}

\hfill\today \\ \\

\baselineskip=28pt
\begin{center}
{\LARGE{\bf A New $p$-Control Chart with Measurement Error Correction}}
\end{center}
\baselineskip=14pt
\vskip 2mm
\begin{center}
Li-Pang Chen\footnote{Email: lchen723@nccu.edu.tw} and Su-Fen Yang\footnote{\baselineskip=10pt Email: yang@mail2.nccu.tw (Corresponding Author). }
\ \\
\ \\
Department of Statistics, National Chengchi University
\end{center}
\bigskip

\vspace{8mm}

\begin{center}
{\Large{\bf Abstract }}
\end{center}
\baselineskip=17pt
{

Control charts are important tools to monitor quality of products. One of useful applications is to monitor the proportion of non-conforming products. However, in practical applications, measurement error is ubiquitous and may occur due to false records or misclassification, which makes the observed proportion different from the underlying true proportion. It is also well-known that ignoring measurement error effects provides biases, and is expected that the resulting control charts may incur wrong detection. In this paper, we study this important problem and propose a valid method to correct for measurement error effects and obtain error-eliminated control chart for the proportion of non-conforming products. In addition, unlike traditional approaches, the corrected EWMA $p$-control chart provides asymmetric control limits and is flexible to handle the data with small sample size. Numerical results are conducted to justify the validity of the corrected EWMA $p$-control chart and verify the necessity of measurement error correction.

}

\vspace{8mm}

\par\vfill\noindent

\par\medskip\noindent
\underline{\bf Short title}:  SPC with misclassification 

\clearpage\pagebreak\newpage
\pagenumbering{arabic}

\newlength{\gnat}
\setlength{\gnat}{22pt}
\baselineskip=\gnat

\clearpage

\section{Introduction}

Statistical process control (SPC) is an important tool in industrial statistics and is useful in monitoring the quality of products. The main interest is to develop control charts based on in-control (IC) statistic, and then use them to monitor and detect out-of-control (OC) process parameters. To do this, one may focus on location process or dispersion process. For example, as summarized in Qiu (2013), many classical methods have been developed under the parametric settings, including Shewhart charts, exponentially weighted moving average (EWMA) charts, and cumulative sum (CUSUM) charts. Without imposing parametric assumptions, distribution free settings have been popular in recent years and relevant approaches have been developed in recent years, such as EWMA mean charts (e.g. Yang 2016), variability monitoring (e.g., Yang and Arnold 2016a; Yang and Wu 2017), and likelihood ratio-based EWMA control charts (e.g., Zou and Tsung 2010). In addition, nonparametric methods have been explored as well. For example, Chen et al. (2022) employed the kernel estimation method to construct the control region based on location and dispersion processes simultaneously. 

In practice, noisy data usually exist and are inevitable in producing products. A typical phenomenon in manufacturer factories is {\em measurement error}, which reflects a fact that observed data are different from underlying true ones that are unobservable. In the presence of continuous random variables, some existing works have examined impacts of measurement error to different kinds of control charts, such as EWMA charts (e.g., Maravelakis et al. 2004; Asif et al. 2020; Nguyen et al. 2020; Tran et al. 2021), Shewhart control charts (e.g., Nguyen et al. 2020; Linna and Woodall 2001),  multivariate process variability (e.g., Huwang and Hung 2007), first-order autoregressive model (e.g., Shongwe et al. 2020), and multivariate control charts (e.g., Linna et al. 2001). However, while those works discussed measurement error effects on control limits, to the best of knowledge, none of them provided suitable strategies to correct for measurement error effects and adjust control limits.

In addition to continuous random variables, sometimes we are interested in the number/proportion of (non-)confirming products. Such a scenario refers binary random variables. Under the assumption of parametric distributions, the strategies of Shewhart charts have been adopted to construct $p$-control charts (e.g., Qiu 2014, Section 3.3). Some relevant extensions have also been explored, such as quasi ARL‐unbiased $p$‐charts based on a heuristic method (e.g., Argoti and Carri\'on‐Garc\'ia 2019), the risk adjustment control charts for categorical random variables (e.g., Sparks 2017), and the general framework of distribution free settings (e.g., Yang and Arnold 2016b; Aslam et al. 2019). Some applications of $p$-control charts have also been discussed. For example, Shu and Wu (2010) applied $p$-control charts to monitor imprecise fraction of conforming items. Bourke (2006) extended the $np$ chart for detecting upward shifts in fraction defective. While the development of $p$-control chart has been widely discussed, however, there are several critical concerns in applications. First, similar to measurement error in continuous random variables, it is possible to encounter measurement error in binary random variables, which refers {\em misclassification} (e.g., Yi 2017, Section 2.6; Chen and Yi 2021). More specifically, when producing products in factories, sometimes they may be falsely detected as confirmed (or non-confirmed) due to imprecise measurement equipment or human-made mistakes. Because of such a misclassification, the observed status is different from what it should be. While measurement error in continuous random variables has been widely discussed, misclassification problem in SPC has been rarely discussed, and strategies for correction of measurement error effects are unavailable. The second concern is the small sample size in each monitoring time. It is known that, by the central limit theorem, the estimated proportion follows normal distributions when the sample size goes to the infinity. However, in the presence of small sample size, the sampling distribution is unknown, and thus, the construction of control charts is unknown.

Due to those concerns, in this paper we focus on SPC for binary random variables subject to misclassification. We particularly focus on phase II scenario where the parameter is assumed to be known. Under small sample sizes, we employ EWMA charts to construct the asymmetric control charts. To deal with measurement error effects, we propose the ``corrected'' proportion of non-confirming products, so that the corresponding control limits can be adjusted accordingly. To assess the performance of the corrected EWMA $p$-control chart, we compare the corrected EWMA $p$-control chart with the naive (uncorrected) EWMA $p$-control chart  that simply adopt error-contaminated random variables. Numerical experiments show that the corrected EWMA $p$-control chart can precisely  detect OC proportion.

The remainder is organized as follows. In Section~\ref{notation}, we introduce the data structure and the EWMA $p$-control chart to monitor non-confirming products. In addition, measurement error to binary random variables is introduced. In Section~\ref{method}, we propose a valid estimation procedure to correct the measurement error effects and adopt the EWMA statistic to construct the asymmetric corrected EWMA $p$-control chart. Moreover, we present in-control average run length (ARL$_0$) and out-of-control average run length (ARL$_1$) to illustrate the out-of-control detection performance. Empirical studies, including simulation results and real data analysis, are provided in Sections~\ref{Sec-Simulation} and~\ref{Sec-RDA}, respectively. We conclude the article with discussions in Section~\ref{Sec-Discussion}. The R programming code for the implementation is available in the Github: \url{https://github.com/lchen723/SPC-ME-R-code.git}. 

\section{Data Structure and Mismeasurement} \label{notation}

In this section, we introduce the construction of the control chart with relevant notation, and discuss the issue of measurement error in binary random variables.

\subsection{Binary Variables and Construction of The {\it p}-Control Chart} \label{p-chart}

Let $T$ denote the monitoring time. For each time $t=1,\cdots, T$, there are $n$ subjects. Let $X_{it}$ for $i=1,\cdots,n$ and $t=1,\cdots,T$ denote the independent and identically distributed (i.i.d.) binary random variable with outcome $0$ or $1$, where $X_{it}=1$ represents a non-conforming product, while $X_{it}=0$ stands for a conforming product. The main interest is to monitor the proportion of non-conforming products for in-control process. 

Let $p_0 \triangleq P(X_{it}=1)$ be the parameter of proportion of non-conforming products for in-control process, and let $q_0 \triangleq 1-p_0 = P(X_{it}=0)$ denote the probability of conforming products for in-control process. Under in-control $n$ subjects for $T$ monitoring times, we define the sample proportion $\widehat{p}_{0,t} \triangleq \frac{1}{n} \sum \limits_{i=1}^n X_{it}$.

To build up the reliable control chart, one may employ the {\em exponentially
weighted moving average} (EWMA) control chart, which is constructed based on a weighted average of in-control sample proportion at the current time point. In addition, as commented in Yang (2016), the EWMA chart is useful in monitoring since it is sensitive in detecting small shifts in process parameters.

Based on the binary random variable and the sample proportion $\widehat{p}_{0,t}$, the in-control EWMA statistic is given by
\begin{eqnarray} \label{CL-p}
\text{EWMA}_{0,t} = \lambda \widehat{p}_{0,t} + (1-\lambda) \text{EWMA}_{0,t-1},
\end{eqnarray}
where $\text{EWMA}_{0,0} = p_0$ with $p_0$ being the in-control proportion, and $\lambda \in (0,1]$ is a smoothing parameter. Moreover, when the monitoring time goes by, the OC non-conforming proportion is non-decreasing and is greater than IC non-conforming proportion. Thus, in applications, it is reasonable to set the lower control limit (LCL) to be zero, and the upper control limit (UCL) is specified as
\begin{eqnarray*}
\text{UCL} = p_0 + L \sqrt{\frac{p_0(1-p_0)\lambda \{1-(1-\lambda)^{2t}\}}{n(2-\lambda)}}.
\end{eqnarray*}
Now, and hereafter, we call this strategy the {\it EWMA $p$-control chart}. 

\subsection{Misclassification}

In practice, variables are often subject to  mismeasurement. That is, instead of collecting unobserved variables $X_{it}$, sometimes we have only the observed or surrogate version of $X_{it}$, denoted by $X_{it}^\ast$. The relationship between unobserved and observed variables $X_{it}, X_{it}^\ast \in \{0,1\}$ can be characterized as $\pi_{kl} = P\big( X_{it}^\ast = k | X_{it} = l \big)$ for $k,l=0,1$. Specifically, $\pi_{00}$ and $\pi_{11}$, which are called {\em classification probability}, indicate that both $X_{it}$ and $X_{it}^\ast$ are either conforming or non-conforming  products; on the other hand, $\pi_{01}$ or $\pi_{10}$ shows that the unobserved variable is opposite to the observed variable which is caused by measurement error. In this case, we call $\pi_{10}$ and $\pi_{01}$ as {\em misclassification probability} (e.g., Carroll et al. 2006; Chen and Yi 2021). To further understand the concept of misclassification, we can think of $X_{it}$ as the {\it true} status of product with $X_{it}=1$ being non-conforming and 0 otherwise, which is unknown. On the other hand, $X_{it}^\ast$ is understood as the {\it observed} status of product that is recorded by factory's staffs. Therefore, $\pi_{01}$ (or $\pi_{10}$) is interpreted as the probability that a non-conforming (or conforming) product is falsely recorded as conforming (or non-conforming).

Let $p_0^\ast = P(X_{it}^\ast = 1)$ and $q_0^\ast = 1-p_0^\ast = P(X_{it}^\ast = 0)$ denote the in-control surrogate version of $p_0$ and $q_0$, respectively. To build up the relationship of $p_0^\ast$ and $p_0$, we apply the technique of law of total probability and obtain
\begin{eqnarray} \label{past}
p_0^\ast = \pi_{11} p_0 + \pi_{10} q_0.
\end{eqnarray}
Similarly, $q_0^\ast$ and $q_0$ can be characterized as
\begin{eqnarray} \label{qast}
q_0^\ast = \pi_{01} p_0 + \pi_{00} q_0.
\end{eqnarray}
The matrix form of (\ref{past}) and (\ref{qast}) is given by
\begin{eqnarray} \label{mis_relation_matrix}
\left(
\begin{array}{c}
p_0^\ast \\
q_0^\ast
\end{array}
\right) = \boldsymbol{\Pi} \left(
\begin{array}{c}
p_0 \\
q_0
\end{array}
\right),
\end{eqnarray}
where $\boldsymbol{\Pi} = \left( 
\begin{array}{cc}
\pi_{11} & \pi_{10} \\
\pi_{01} & \pi_{00}
\end{array}
\right)$ is the $2 \times 2$ (mis)classification matrix and the sum of elements in each row is equal to one, i.e., $\pi_{11} + \pi_{10} = 1$ and $\pi_{01} + \pi_{00} = 1$. Following the similar discussion in Chen and Yi (2021), we assume that $\boldsymbol{\Pi}$ has the spectral decomposition $\boldsymbol{\Pi} = \boldsymbol{\Omega} \mathbf{D} \boldsymbol{\Omega}^{-1}$, where $\mathbf{D}$ is the diagonal matrix with diagonal elements being the eigenvalues of $\boldsymbol{\Pi}$, and $\boldsymbol{\Omega}$ is the corresponding matrix of eigenvectors.

Since our main target is to monitor $p_0$, however, from the equality (\ref{mis_relation_matrix}), we observe that the surrogate version $p_0^\ast$ is no longer equal to $p_0$ if misclassification occurs, i.e., $\pi_{10} \neq 0$ or $\pi_{01} \neq 0$. In particular, when $\pi_{10}$ or $\pi_{01}$ become larger, $p_0^\ast$ and $q_0^\ast$ are different from $p_0$ and $q_0$. Therefore, with the availability of $X_{it}^\ast$, the sample proportion $\widehat{p}_{0,t}^\ast = \frac{1}{n} \sum \limits_{i=1}^n X_{it}^\ast$ and $\widehat{q}_{0,t}^\ast = 1-\widehat{p}_{0,t}^\ast$ have biases for $p_0$ and $q_0$, and the corresponding control limits, determined by (\ref{CL-p}) with $\widehat{p}_{0,t}$ replaced by $\widehat{p}_{0,t}^\ast$, may incur wrong detection.

\subsection{Determination of Classification Matrix}

While the probability of observed non-conforming or conforming products can be expressed as (\ref{mis_relation_matrix}), the (mis)classification matrix $\boldsymbol{\Pi}$ is usually unknown because of involvement of the unobserved variable $X_{it}$. Therefore, to develop the method,  we consider sensitivity analyses (e.g., Chen 2020; Chen and Yi 2021), whose  purpose is to specify different values of $\boldsymbol{\Pi}$ to understand how mismeasurement effects may affect inference results and control charts, and it is usually employed when  additional information is unavailable.

Since the specification of $\boldsymbol{\Pi}$ may be not unique, in this paper we consider the relative ratio (RR) to specify misclassification probabilities. Specifically, let the relative ratio under $X_{it}=1$ be
\begin{eqnarray*}
\text{RR}_1 = \frac{\pi_{11}}{1-\pi_{11}},
\end{eqnarray*}
where $\text{RR}_1$ is in an interval $[0,\infty)$. A larger $\text{RR}_1$ indicates a higher probability of correct classification. Then given $\text{RR}_1$, $\pi_{11}$ and $\pi_{01}$ can be derived as $\pi_{11} = \frac{\text{RR}_1}{1+\text{RR}_1}$ and $\pi_{01} = \frac{1}{1+\text{RR}_1}$, respectively. Similarly, under $X_{it}=0$, the relative ratio is given by
\begin{eqnarray*}
\text{RR}_0 = \frac{\pi_{00}}{1-\pi_{00}}.
\end{eqnarray*}
Then for a specified $\text{RR}_0$, $\pi_{00}$ and $\pi_{10}$ are given by $\pi_{00} = \frac{\text{RR}_0}{1+\text{RR}_0}$ and $\pi_{10} = \frac{1}{1+\text{RR}_0}$, respectively. Therefore, $\boldsymbol{\Pi}$ can be obtained.

\section{Methodology} \label{method}


\subsection{Corrected Control Limits under Expectation} \label{correct-p}

Inspired by (\ref{mis_relation_matrix}), an intuitive approach is to take the inverse matrix $\boldsymbol{\Pi}^{-1}$ to both sides of (\ref{mis_relation_matrix}), yielding 
\begin{eqnarray} \label{mis_relation_matrix_correct}
\boldsymbol{\Pi}^{-1} \left(
\begin{array}{c}
p_0^\ast \\
q_0^\ast
\end{array}
\right) = \left(
\begin{array}{c}
p_0 \\
q_0
\end{array}
\right).
\end{eqnarray}
It suggests that $\boldsymbol{\Pi}^{-1}$ is regarded as the term to correct for error effects in $p_0^\ast$ and $q_0^\ast$, and the left-hand-side of (\ref{mis_relation_matrix_correct}) can reflect $p_0$ and $q_0$.

Specifically, from (\ref{mis_relation_matrix_correct}), $p_0$ can be expressed by
\begin{eqnarray} \label{key-method-1}
p_0 
&=& \frac{\pi_{00} p_0^\ast - \pi_{10} q_0^\ast }{\pi_{11} \pi_{00} - \pi_{10} \pi_{01}} \nonumber \\
&=&  \frac{(1-\pi_{10}) p_0^\ast - \pi_{10} q_0^\ast }{(1-\pi_{10}) (1-\pi_{01}) - \pi_{10} \pi_{01}} \nonumber \\
&=&  \frac{p_0^\ast - \pi_{10} }{1-\pi_{10} -\pi_{01}}, 
\end{eqnarray}
where the first equality is due to the manipulation of inverse matrix, the second equality is due to the property of $\boldsymbol{\Pi}$, and the last step comes from the fact $p_0^\ast + q_0^\ast = 1$. Consequently, (\ref{key-method-1}) suggests that the ``corrected'' proportion of non-conforming products, denoted as $p_0^{\ast\ast}$, is defined as the right-hand-side of (\ref{key-method-1}).
That is,
\begin{eqnarray} \label{est-p-mathod-1}
p_0^{\ast\ast} \triangleq \frac{p_0^\ast - \pi_{10} }{1-\pi_{10} -\pi_{01}}.
\end{eqnarray}
Thus, it suggests the {\it corrected} random variable be
\begin{eqnarray*}
X_{it}^{\ast\ast} \triangleq \frac{X_{it}^\ast - \pi_{10} }{1-\pi_{10} -\pi_{01}}.
\end{eqnarray*}
Thus, the corrected sample proportion is denoted as $\widehat{p}_{0,t}^{\ast\ast} = \frac{1}{n} \sum \limits_{i=1}^n X_{it}^{\ast\ast}$. 

\subsection{An Error Corrected Asymmetric EWMA $p$-Control Chart}

As discussed in Section~\ref{p-chart}, we can take the EWMA as the monitoring statistic and then develop the control chart. To address this, we apply (\ref{CL-p}) with $\widehat{p}_{0,t}$ replaced by $\widehat{p}_{0,t}^{\ast\ast}$ to obtain the corrected EWMA statistics, which is given by
\begin{eqnarray} \label{EWMA-p-chart}
\text{EWMA}_{0,t}^{\ast\ast} = \lambda \widehat{p}_{0,t}^{\ast\ast} + (1-\lambda) \text{EWMA}_{0,t-1}^{\ast\ast}
\end{eqnarray}
for $t=1,\cdots,T$, where $\lambda \in (0,1]$ is a smoothing parameter.

To construct the control chart, we require the expectation and variance of (\ref{EWMA-p-chart}) as stated in the following theorem.

\begin{theorem} \label{Thm-EWMA}
Under the monitoring statistic (\ref{EWMA-p-chart}), we have
\begin{itemize}
    \item[(a)] $E\left\{ \text{EWMA}_{0,t}^{\ast\ast} \right\} = p_0  $;
    
    \item[(b)] $\text{var}\left\{ \text{EWMA}_{0,t}^{\ast\ast} \right\} = \frac{p_0^\ast(1-p_0^\ast)\lambda \{1-(1-\lambda)^{2t}\}}{n(1 - \pi_{10} - \pi_{01})^2(2-\lambda)}   $.
\end{itemize}
\end{theorem}

Under Theorem~\ref{Thm-EWMA}, the corrected UCL and LCL are given by
\begin{eqnarray} \label{correct-CL}
\text{UCL}^{\ast\ast} = p_0^{\ast\ast} + L^{\ast\ast} \sqrt{\frac{p_0^\ast(1-p_0^\ast)\lambda \{1-(1-\lambda)^{2t}\}}{n(1 - \pi_{10} - \pi_{01})^2(2-\lambda)}}, \ \ \text{and} \ \ \text{LCL}^{\ast\ast}=0.
\end{eqnarray}
We call (\ref{correct-CL}) the {\it corrected} EWMA $p$-control chart.

\begin{remark}
\
\begin{enumerate}
    \item If the interest is $np$ charts, then simply multiplying $n$ to (\ref{correct-CL}) yields the desired result.

\item If sample sizes in different subgroup are different from each other, then the sample size $n$ in (\ref{correct-CL}) can be replaced by $n_k$ for the $k$th subgroup.

\end{enumerate}

\end{remark}

\begin{remark} (Naive EWMA statistic and $p$-control chart) \label{Naive_EWMA}

Under the observed random variable \textnormal{without} error correction, we can also construct the EWMA chart. Specifically, let
\begin{eqnarray} \label{EWMA-p-chart-naive}
\text{EWMA}_{0,t}^\ast = \lambda \widehat{p}_{0,t}^\ast + (1-\lambda) \text{EWMA}_{0,t-1}^\ast
\end{eqnarray}
denote the naive EWMA statistic, and we define
\begin{eqnarray} \label{Naive_chart}
\text{UCL}^\ast = p_0^\ast + L^\ast \sqrt{\frac{p_0^\ast(1-p_0^\ast)\lambda \{1-(1-\lambda)^{2t}\}}{n(2-\lambda)}}, \ \ \text{LCL}^\ast=0
\end{eqnarray}
as the naive EWMA $p$-control chart.

\end{remark}

\subsection{Examination of ARL}

In the framework of statistical process control, to assess the performance of $p$-control charts, we usually examine the average run length (ARL). 

Let $\text{ARL}_0$ denote the in-control (IC) ARL, which reflects the mean of the run length when the process is IC. In addition, let $\text{ARL}_1$ be the out-of-control (OC) ARL, which gives the average number of samples collected from the time of shift occurrence to the time of signal. 

The goal is to use (\ref{correct-CL}) to compute ARL$_0$ and ARL$_1$. Unlike the Shewhart chart where ARL$_0$ and ARL$_1$ can reflect Type I and II errors, respectively, the challenge is that the EWMA statistic has no such a property since it is dependent. In addition, the coefficient of control limit is usually unknown and the control limit is asymmetric. Therefore, to deal with those issues, we first fix ARL$_0$ at a given level, and then the calculate the coefficient of control limit $L^{\ast\ast}$. After that, based on the constructed EWMA chart, we compute ARL$_1$. Detailed computational algorithms are placed in the following two subsections, and similar strategy can be applied to the naive EWMA $p$-control chart (\ref{Naive_chart}).



\subsubsection{Determination of Critical Values and In-Control ARL} \label{compute-ARL0}

We first fix $\text{ARL}_0$ whose typical choices include 200 or 370. In addition, under the error-prone data, we can use (\ref{mis_relation_matrix}) to calculate $p_0^\ast$. To address measurement error effects, we employ (\ref{est-p-mathod-1}) to derive the corrected IC proportion $p_0^{\ast\ast}$.

Given a range $(a,b)$ with  user-specific values $a=0.01$ and $b=10$ for $L^{\ast\ast}$, we aim to find the optimal value of  $L^{\ast\ast}$ that satisfies a given ARL$_0$. The strategy to determine $L^{\ast\ast}$ is the Monte Carlo method with repetition $M$. Under the $m$th Monte Carlo step, we evaluate (\ref{EWMA-p-chart}) and $\text{UCL}^{\ast\ast}$ in (\ref{correct-CL}) for every $t=1,2,\cdots,T$ based on a proportion determined by IC samples. Let $t_{0,m}$ denote a run length, which is defined as a value $t$ such that $\text{EWMA}_{0,t}^{\ast\ast} \geq \text{UCL}^{\ast\ast}$. Finally, under $M$ repetitions in the Monte Carlo procedure, the estimated average run length under IC samples is given by $\widehat{\text{ARL}}_0 \triangleq \frac{1}{M} \sum \limits_{m=1}^M t_{0,m}$, and thus, $L^{\ast\ast} \in (a , b)$ can be determined by minimizing $\Big| \text{ARL}_0 - \widehat{\text{ARL}}_0 \Big| < 1$. The pseudo-code of computational procedure for critical value $L^{\ast\ast}$ and ARL$_0$ is presented in Algorithm~\ref{ARL0-IC}.

\subsubsection{Determination of Out-of Control ARL}

Let $p_1$ denote the OC process proportion that is determined by re-scaled proportion $p_0$. Moreover, in the presence of measurement error, we have the misclassified OC proportion $p_1^\ast$, which can be characterized with respect to $p_1$ by the similar relationship of  (\ref{mis_relation_matrix}). To correct for measurement error effects and recover misclassified proportion to the true one, we again employ (\ref{est-p-mathod-1}) to define the adjusted OC proportion, and denote it as $p_1^{\ast\ast}$.

After adjusting the OC proportion, the next goal is to examine the constructed control chart and compute ARL$_1$. Following the similar strategy in Section~\ref{compute-ARL0}, we employ the Monte Carlo method to derive ARL$_1$ based on the constructed EWMA control chart. For the $m$th step in the Monte Carlo procedure, we use (\ref{EWMA-p-chart}) to compute EWMA based on OC proportion for each monitor time $t$. After that, let $t_{1,m}$ denote an OC run length, which is given by a value $t$ that satisfies $\text{EWMA}_{1,t}^{\ast\ast} \geq \text{UCL}^{\ast\ast}$ with $\text{UCL}^{\ast\ast}$ being determined in Algorithm~\ref{ARL0-IC}. Consequently, under $M$ repetitions in the Monte Carlo procedure, the estimated average run length under OC samples is given by $\widehat{\text{ARL}}_1 = \frac{1}{M} \sum \limits_{m=1}^M t_{1,m}$. The pseudo-code of computation of ARL$_1$ is presented in Algorithm~\ref{AR$L_1$-OC}.

\section{Simulation Studies} \label{Sec-Simulation}

\subsection{Simulation Setup} \label{Sec-Sim-Set}

Let $n$ denote the sample size in each monitoring time, where $n=5,10,15,$ and $20$. Let $T$ be the monitoring time that is set as $T = 5000$. For $i=1,\cdots,n$ and $t=1,\cdots,T$, let the true IC data $X_{it}$ be generated from the Bernoulli distribution with the IC probability $p_0$ that is specified as 0.05, 0.1, 0.15, 0.2, 0.25, 0.3, 0.35, 0.4, 0.45, 0.5. Moreover, for the OC samples, the proportion $p_1$ is defined as $p_1 \triangleq (1+\delta) p_0$ with $\delta$ being specified as 0.1 and 0.2.

For the misclassification model (\ref{mis_relation_matrix}), we specify $\pi_{00} = \pi_{11} = \pi$ and $\pi_{10} = \pi_{01} = 1-\pi$ with $\pi =0.95$ and $0.99$, yielding reliability ratios $\text{RR}_1 = \text{RR}_0 =0.474,0.487, 0.498$, respectively. Based on the misclassification model (\ref{mis_relation_matrix}), we generate the error-prone binary random variable $X^\ast$. We denote $p_0^\ast$ and $p_1^\ast$ as the corresponding error-prone IC and OC probabilities, respectively.

To implement the corrected EWMA $p$-control chart, we adopt (\ref{est-p-mathod-1}) to calculate the corrected probability $p_0^{\ast\ast}$, and then employ (\ref{EWMA-p-chart}) to construct the EWMA statistics with smoothing parameters being $\lambda = 0.05$ or $0.2$. Finally, 10000 simulations are run for all settings.

\subsection{Simulation Results}

To assess the OC detection performance of the corrected EWMA $p$-control chart, we also investigate the naive method in Remark~\ref{Naive_EWMA} and (\ref{CL-p}) by assuming the existence of the true $X$. While the common choice of ARL$_0$ can be 370 or 200, here we only explore 370 because the result under ARL$_0=200$ has the similar pattern to that based on ARL$_0=370$.

We first present numerical results for coefficients of the control limit and UCLs based on given $\pi$ and $\lambda$ in Tables~\ref{Sim_370_0.05_0.95}-\ref{Sim_370_0.2_0.99}. We observe that UCL decreases when $n$ becomes large regardless of the methods. In the presence of misclassification, we observe that UCL$^\ast$ increases when misclassification is severe (i.e., $\pi$ becomes small); on the contrary, it is interesting to see that values of UCL$^{\ast\ast}$ determined by the corrected EWMA $p$-control chart keep remained and are the same as UCL based on the true $X$. In addition, values of UCL$^\ast$ are greater than others, suggesting wider control limits determined by error-prone variables. On the other hand, with $\pi$ fixed, values of UCL will increase when $\lambda$ is increasing regardless of the implementation of methods. Regarding the results of coefficients of the control limit $L$, we observe that $L^{\ast\ast}$ is smaller than $L^\ast$ for most settings, and values of $L^{\ast\ast}$ have the same decreasing pattern as values of $\pi$ and $\lambda$ do.

After obtaining estimated control limits, we next assess the estimation results of ARL$_1$. Numerical results under ARL$_0=370$ are placed in Tables~\ref{ARL1_370_0.05_0.95}-\ref{ARL1_370_0.2_0.99}, where ARL$_1$ with superscripts $\ast$ and $\ast\ast$ represent the naive and proposed methods, respectively, and the results for the true $X$ are recorded as ARL$_1$ without the superscript. We observe that all values decease when $n$, $p_1$, or $\delta$ increase. In general, we can see that values of ARL$_1^\ast$ are always greater than ARL$_1$ and ARL$_1^{\ast\ast}$, and differences become large as $\delta$ and $n$ are small. It shows that the naive method is unsatisfactory to detect OC proportion, because smaller values of ARL$_1$ have better performance to detect OC (e.g., Maravelakis et al. 2004). On the other hand, values of ARL$_1^{\ast\ast}$ are close to ARL$_1$ under the true $X$. These results suggest that (1) the necessity of measurement error correction, (2) the correction of measurement error effects recover the control limit to that determined by the true $X$, and makes the detection of OC be the same as that of true $X$, (3) the corrected EWMA $p$-control chart is successful to handle the scenario of small sample size and is robust in detecting OC proportion.

\section{Application to Real Data Analysis} \label{Sec-RDA}

In this section, we apply the corrected EWMA $p$-control chart to the orange juice data, which can be found in the R package \texttt{qcr}.

The primary interest of this study is the production of 6-oz cardboard cans that frozen orange juice concentrate is packed in. These cans are formed on a machine by spinning them from cardboard stock and attaching a metal bottom panel. A can is then inspected to determine whether, when filled, the liquid could possible leak either on the side seam or around the bottom joint. If this situation occurs, then a can is considered non-conforming. The data were collected as 30 samples of 50 cans each at half-hour intervals over a three-shift period in which the machine was in continuous operation. From sample 15 used, a new batch of cardboard stock was punt into production. Sample 23 was obtained when an inexperienced operator was temporarily assigned to the machine. After the first 30 samples, a machine adjustment was made. Then further 24 samples were taken from the process. Therefore, according to the data description, we take the IC monitoring time as $T = 24$ and assign the OC monitoring time as 30. For each IC monitoring time, the sample size is $n=50$.

In applications, measurement error is ubiquitous in process monitoring. In addition,  Pendrill (2014) and Puydarrieux et al. (2019) pointed out that risks arising from measurement are main concern in conformity assessment. In other words, from the orange juice data, it is possible that factory's staffs falsely record non-conforming (or conforming) cans to be conforming (or non-conforming). As a result, it is reasonable to assume that the collected data are subject to measurement error, and the corresponding proportion is given by $p_0^\ast = 0.111$.

Since measurement error correction is crucial, we now adopt (\ref{mis_relation_matrix}) to to characterize measurement error and employ the corrected EWMA $p$-control chart (\ref{est-p-mathod-1}) to correct for measurement error effects. Since  $\boldsymbol{\Pi}$ in (\ref{mis_relation_matrix}) is unknown and additional information is unavailable, we employ sensitivity analyses to address different levels of measurement error effects. Specifically, we follow Section~\ref{Sec-Sim-Set} to specify $\pi_{00} = \pi_{11} = \pi$ and $\pi_{10} = \pi_{01} = 1-\pi$ with $\pi = 0.95$ and $0.99$, which yield $p_0^{\ast\ast} = 0.068$ and $0.103$, respectively.

Now, for each fixed $\pi=0.95$ or $0.99$ and $\lambda = 0.05$ or $0.2$, we adopt (\ref{EWMA-p-chart}) and (\ref{EWMA-p-chart-naive}) to compute the EWMA statistics, and apply Algorithm~\ref{ARL0-IC} to compute $L$ and UCL based on a given ARL$_0$, yielding the corrected or naive EWMA $p$-control charts, respectively. For the choice of ARL$_0$, we simply use 370 for exploration, and ARL$_0=200$ gives the similar result.

\begin{figure}[!ht]
\centering
\begin{tabular}{ccc}
    \includegraphics[width=.8\textwidth]{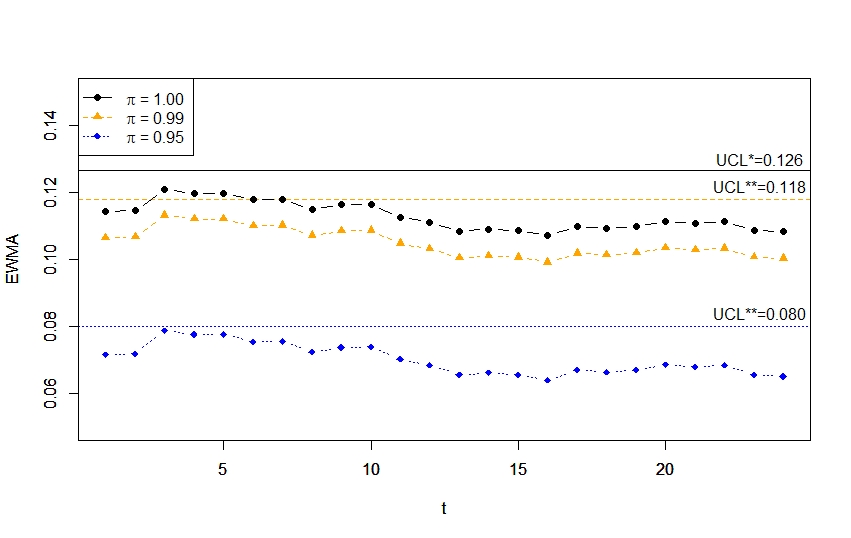}
\end{tabular}
\caption{The naive and corrected EWMA $p$-control charts under $\lambda=0.05$ and IC samples.}
\label{IC_ARL370_lambda005}
\end{figure}

\begin{figure}[!ht]
\centering
\begin{tabular}{ccc}
    \includegraphics[width=.8\textwidth]{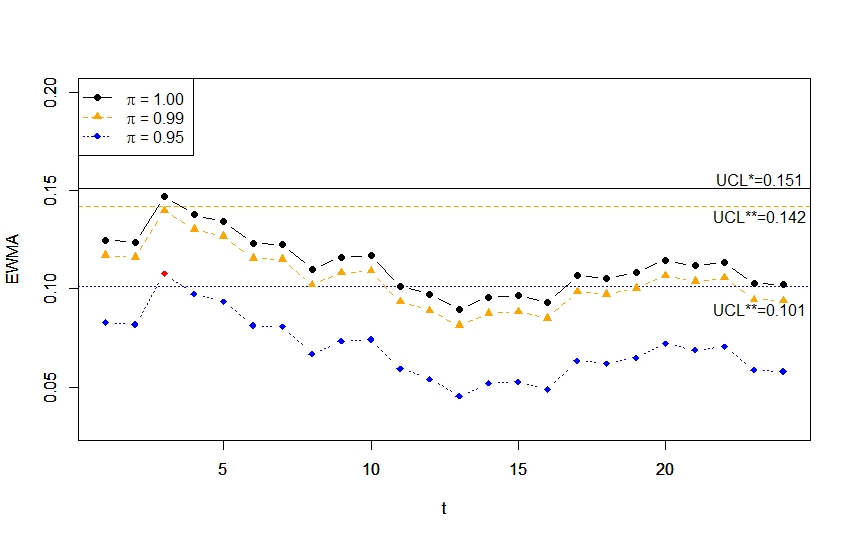}
\end{tabular} 
\caption{The naive and corrected EWMA $p$-control charts under $\lambda=0.20$ and IC samples.}  
\label{IC_ARL370_lambda02}
\end{figure}

We first report the estimation results for $L$ and UCL based on the naive and corrected EWMA $p$-control charts in Table~\ref{RDA}. We observe that coefficients $L$ and UCL values based on the naive EWMA $p$-control chart are larger than those based on the corrected EWMA $p$-control chart regardless of choices of $\lambda$, which indicates that the naive EWMA $p$-control chart gives wider control limits than those given by the corrected EWMA $p$-control chart. In addition, from the corrected EWMA $p$-control chart, we can see that both $L$ and UCL are decreasing when $\pi$ becomes small, which reflects a phenomenon that the corrected control limits may be implicitly affected by the proportion of misclassification. Moreover, we also display naive and corrected EWMA $p$-control charts based on $\lambda=0.05$ and $0.2$ in Figures~\ref{IC_ARL370_lambda005} and~\ref{IC_ARL370_lambda02}, respectively, for visualization. We can see that all monitoring points are under the upper control limit, except for the third point that is known as the false alarm based on $\pi=0.95$ in Figure~\ref{IC_ARL370_lambda02} it may show that severe misclassification makes detection more sensitive.

\begin{figure}[!ht]
\centering
\begin{tabular}{ccc}
    \includegraphics[width=.8\textwidth]{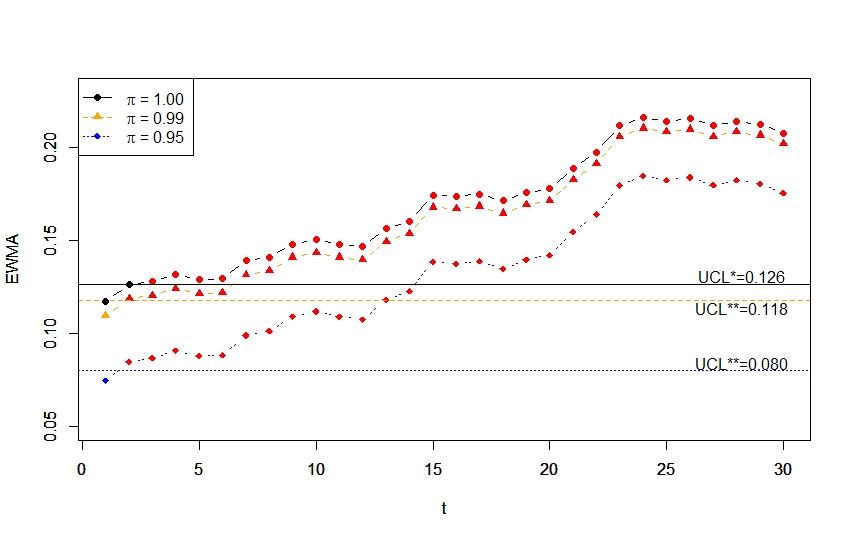}
\end{tabular}
\caption{The naive and corrected EWMA $p$-control charts under $\lambda=0.05$ and OC samples. Red points are OC detection.}
\label{OC_ARL370_lambda005}
\end{figure}

\begin{figure}[!ht]
\centering
\begin{tabular}{ccc}
    \includegraphics[width=.8\textwidth]{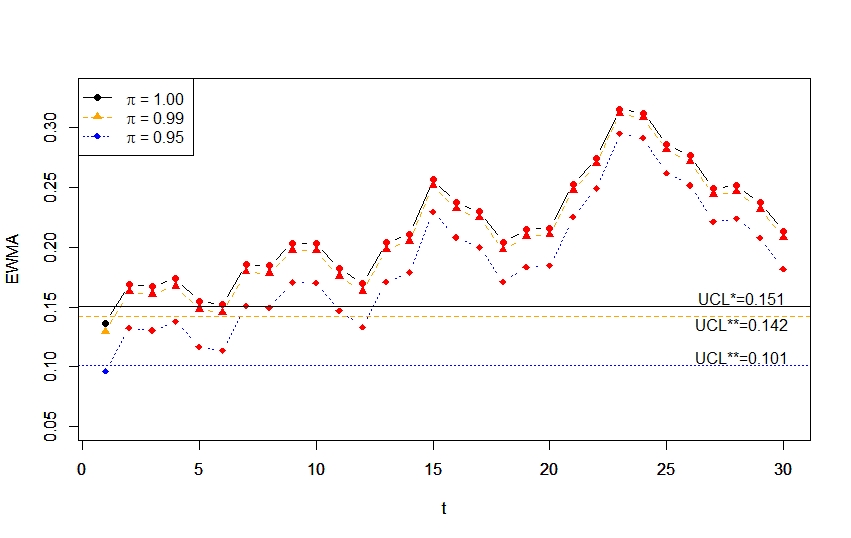}
\end{tabular} 
\caption{The naive and corrected EWMA $p$-control charts under $\lambda=0.20$ and OC samples. Red points are OC detection.}  
\label{OC_ARL370_lambda02}
\end{figure}

Finally, based on the developed naive or corrected EWMA $p$-control charts, we further detect OC process parameters, and the associated control charts under $\lambda=0.05$ or $0.2$ and $\pi=0.95$ or $0.99$ are displayed in Figures~\ref{OC_ARL370_lambda005} and~\ref{OC_ARL370_lambda02}, respectively. We observe that detected points under uncorrected or corrected EWMA $p$-control charts have similar pattern, and values of EWMA statistic become small as $\pi$ is decreasing. In addition, the naive EWMA $p$-control chart seems to be sensitive to detect OC with the choice of $\lambda$. In particular, when $\lambda=0.05$, the second point of the naive EWMA $p$-control chart in Figure~\ref{OC_ARL370_lambda005} is not detected as OC, while the corrected EWMA $p$-control chart can successfully detect OC regardless of the choices of $\lambda$ or $\pi$.


\section{Discussion} \label{Sec-Discussion}

Statistical process control has been an important tool to monitor products. Some methods have been developed to construct control charts for binary and/or continuous random variables. In applications, however, measurement error exists due to imprecise operation systems or human-made mistakes, and ignoring measurement error effects may cause wrong detection. While many research works have been available to discuss measurement error effects on continuous random variables, however, to the best of our knowledge, little attention focused on binary random variables. Moreover, none of them provided suitable strategies to adjust error effects. In this paper, we consider binary random variables that can be used to describe non-conforming product, and we allow binary random variables to be contaminated by misclassification. We proposed the corrected proportion to adjust for measurement error effects, and then employ the corrected EWMA $p$-control chart to obtain reliable and asymmetric control chart under small sample size. Numerical results based on different setting verify the validity of the corrected EWMA $p$-control chart.

One of the concerns in misclassification model (\ref{mis_relation_matrix}) is the determination of $\boldsymbol{\Pi}$. In our development, we employ sensitivity analyses to explore the impact of measurement error effects by examining different levels of misclassification probabilities. In the framework of measurement error analysis, $\boldsymbol{\Pi}$ can be estimated if auxiliary information is available. One of typical information is external validation data. Specifically, suppose that $\mathcal{M}$ with $|\mathcal{M}|=n$ is the subject set for the main study containing measurements $\left\{ X_{it}^\ast : i \in \mathcal{M}, t=1,\cdots,T \right\}$ and let $\mathcal{V}$ with $|\mathcal{V}|=m$ denote the subject set for the external validation study containing measurements $\left\{ \left( X_{it}, X_{it}^\ast \right): i \in \mathcal{V}, t=1,\cdots,T \right\}$, where $\mathcal{M}$ and $\mathcal{V}$ do not overlap. Assume that the main study and the validation study share the same model (\ref{mis_relation_matrix}). With the availability of external validation data $\mathcal{V}$, we have a $2 \times 2$ confusion table. Then the probability $\pi_{kl}$ for $k,l \in \{0,1\}$ can be estimated by 
\begin{eqnarray*}
\widehat{\pi}_{kl} = \frac{\text{number of }\left\{ X_{it}^\ast = k \ \text{and} \ X_{it} = l \right\} \text{ for all } i \in \mathcal{V} \text{ and } t=1,\cdots,T}{\text{number of }\left\{ X_{it} = l \right\} \text{ for all } i \in \mathcal{V} \text{ and } t=1,\cdots,T}.
\end{eqnarray*} 
Therefore, $\boldsymbol{\Pi}$ can be estimated by $\widehat{\boldsymbol{\Pi}} = \big[\widehat{\pi}_{kl} \big]_{k,l\in\{0,1\}}$.

There are some possible extensions for this project. First of all, in addition to the current strategy in Section~\ref{correct-p} to address measurement error effects, in the framework of measurement error analysis, simulation and extrapolation (SIMEX) method  (e.g., Chen 2020) is also a valid tool to measurement error effects. It is a worth exploration in depth because no relevant work has been available for SPC. Moreover, the corrected EWMA $p$-control chart can be naturally extended to other settings. For example, as developed by Yang and Arnold (2016a, 2016b) and Yang (2017), to address distribution free continuous random variables and build up control chart, a common approach to translate the continuous random variables to binary ones. In the presence of measurement error, measurement error effects may affect the translated random variables and associated proportion. Therefore, the corrected EWMA $p$-control chart and the corrected proportion can be employed to deal with this issue. The other interesting is about the profile monitoring (e.g., Qiu 2014, Chapter 10), which aims to take auxiliary information as the covariates, and then build up a regression model whose responses are the main interest to be monitored. In the case of monitoring confirming product, we may consider logistic regression models, and the similar strategy in Section~\ref{correct-p} can be adopted to address this concern. Detailed and deep explorations can be our next research projects in the future.



\clearpage

\begin{algorithm}[H]
\SetAlgoLined
 {\bf Step 1:} Given in-control $p_0$ and $\boldsymbol{\Pi}$, $p_0^\ast$ is calculated by formula (\ref{mis_relation_matrix})\; 
 {\bf Step 2:} Set $\lambda$, $n$, and a value of ARL$_0$\;
 
 {\bf Step 3:} Set $a < L^{\ast\ast} < b$ with $a=0.01$ and $b=10$ (say)\;

 {\bf Step 4:} Monte Carlo procedure:\\
 \For{step $(m+1)$ with $m=1,2,\cdots,M$ and set $M=10001$ (say)}{
 {\bf Step 4.1:} Let $\text{EWMA}_{0,0}^{\ast\ast} = p_0^{\ast\ast}$ and $t=1$\;
  {\bf Step 4.2:} Based on the observed and misclassified $X_{it}^\ast$, calculate $\widehat{p}_{0,t}^{\ast\ast} \triangleq \frac{1}{n} \sum \limits_{i=1}^n \frac{X_{it}^\ast - \pi_{10} }{1-\pi_{10} -\pi_{01}}$ \;
 \begin{itemize}
     \item If $t=1$, $\text{EWMA}_{0,1}^{\ast\ast} = \lambda \widehat{p}_{0,1}^{\ast\ast} + (1-\lambda) \text{EWMA}_{0,0}^{\ast\ast}$;
     
     \item If $t\neq 1$, $\text{EWMA}_{0,t}^{\ast\ast} = \lambda \widehat{p}_{0,t}^{\ast\ast} + (1-\lambda) \text{EWMA}_{0,t-1}^{\ast\ast}$.
 \end{itemize}

  {\bf Step 4.3:} Set UCL$^{\ast\ast}$ and LCL$^{\ast\ast}$ as values in (\ref{correct-CL}) \;
 \begin{itemize}
     \item If  $\text{EWMA}_{0,t}^{\ast\ast} \geq \text{UCL}^{\ast\ast}$, then take $t_{0,m} \triangleq t$ as a run length and $m \leftarrow m+1$. 
     
     Go to Step 1;
     
     \item If  $0<\text{EWMA}_{0,t}^{\ast\ast} \leq \text{UCL}^{\ast\ast}$, then $t \leftarrow t+1$. Go to Step 2.
 \end{itemize}
 }

\begin{description}
\item[\bf Step 5:] Calculate $\frac{1}{M} \sum \limits_{m=1}^M t_{0,m}$ and take it as the estimate of the nominal ARL$_0$, which is denoted as $\widehat{\text{ARL}}_0$. Finally, determine $L^{\ast\ast}$ by minimizing $\Big| \text{ARL}_0 - \widehat{\text{ARL}}_0 \Big| < 1$ subject to $a < L^{\ast\ast} < b$. 
\end{description}
 \caption{Monte Carlo simulation to find coefficients of the control limit and ARL$_0$} \label{ARL0-IC}
\end{algorithm}

\clearpage

\begin{algorithm}[H]
\SetAlgoLined
\begin{description}
 \item[\bf Step 1:] Given out-of-control $p_1^\ast$, $\lambda$, $n$, a nominal ARL$_0$, and UCL$^{\ast\ast}$\; 
 
 \item[\bf Step 2:] Based on $X_{it}^\ast$ in OC, calculate $\widehat{p}_{1,t}^{\ast\ast} \triangleq \frac{1}{n} \sum \limits_{i=1}^n \frac{X_{it}^\ast - \pi_{10} }{1-\pi_{10} -\pi_{01}}$\;
 \end{description}
{\bf Step 3:} Monte Carlo procedure:

 \For{step $(m+1)$ with $m=1,2,\cdots,M$ and set $M=10001$ (say)}{
 {\bf Step 3.1:} Let $\text{EWMA}_{1,0}^{\ast\ast} = p_0^{\ast\ast}$ and $t=1$\;
  {\bf Step 3.2:} calculate out-of-control statistics, denoted $ \text{EWMA}_{1,t}^{\ast\ast}$:
 \begin{itemize}
     \item If $t=1$, $\text{EWMA}_{1,1}^{\ast\ast} = \lambda \widehat{p}_{1,1}^{\ast\ast} + (1-\lambda) \text{EWMA}_{1,0}^{\ast\ast}$;
     
     \item If $t\neq 1$, $\text{EWMA}_{1,t}^{\ast\ast} = \lambda \widehat{p}_{1,t}^{\ast\ast} + (1-\lambda) \text{EWMA}_{1,t-1}^{\ast\ast}$.
 \end{itemize}

  {\bf Step 3.3:} Plot $\text{EWMA}_{1,t}^{\ast\ast}$ in the chart.
 \begin{itemize}
     \item If  $\text{EWMA}_{1,t}^{\ast\ast} \geq \text{UCL}^{\ast\ast}$, then take $t_{1,m} \triangleq t$ as a run length and $m \leftarrow m+1$. 
     
     Go to Step 3.1;
     
     \item If  $\text{EWMA}_{1,t}^{\ast\ast} \leq \text{UCL}^{\ast\ast}$, then $t \leftarrow t+1$. Go to Step 3.2.
 \end{itemize}
 }

{\bf Step 4:} Calculate the estimated ARL$_1$, which is given by $\widehat{\text{ARL}}_1 = \frac{1}{M} \sum \limits_{m=1}^M t_{1,m}$.

 \caption{Calculation for ARL$_1$} \label{AR$L_1$-OC}
\end{algorithm}

\clearpage

\begin{table}
\caption{Simulation results ARL$_0=370$, $\lambda = 0.05$, and $\pi=0.95$.} \label{Sim_370_0.05_0.95}
\begin{tabular}{lccccccccccc}
 \hline
\multicolumn{1}{c}{}   & $p_0$   & 0.050  & 0.100   & 0.150  & 0.200   & 0.250  & 0.300   & 0.350  & 0.400   & 0.450  & 0.500   \\
                       & $p_0^\ast$  & 0.095 & 0.140  & 0.185 & 0.230  & 0.275 & 0.320  & 0.365 & 0.410  & 0.455 & 0.500   \\
\multicolumn{1}{c}{$n$}  & $p_0^{\ast\ast}$   & 0.050  & 0.100   & 0.150  & 0.200   & 0.250  & 0.300   & 0.350  & 0.400   & 0.450  & 0.500   \\
 \hline
\multicolumn{1}{c}{5}  & $L$   & 2.463 & 2.355 & 2.290 & 2.281 & 2.260 & 2.242 & 2.222 & 2.205 & 2.195 & 2.185 \\
                       & UCL & 0.088 & 0.151 & 0.209 & 0.265 & 0.320 & 0.374 & 0.426 & 0.477 & 0.528 & 0.578 \\
                       & $L^\ast$   & 2.363 & 2.311 & 2.305 & 2.263 & 2.261 & 2.236 & 2.215 & 2.205 & 2.194 & 2.184 \\
                       & UCL$^\ast$ & 0.145 & 0.197 & 0.249 & 0.298 & 0.347 & 0.395 & 0.441 & 0.488 & 0.533 & 0.578 \\
                       & $L^{\ast\ast}$   & 1.645 & 1.833 & 1.902 & 1.955 & 1.975 & 1.976 & 1.980 & 1.980 & 1.982 & 1.966 \\
                       & UCL$^{\ast\ast}$ & 0.088 & 0.151 & 0.209 & 0.265 & 0.320 & 0.373 & 0.426 & 0.477 & 0.529 & 0.578 \\
 \hline
\multicolumn{1}{c}{10} & $L$   & 2.374 & 2.295 & 2.270 & 2.258 & 2.230 & 2.221 & 2.211 & 2.202 & 2.185 & 2.182 \\
                       & UCL & 0.076 & 0.135 & 0.191 & 0.246 & 0.299 & 0.352 & 0.403 & 0.455 & 0.505 & 0.555 \\
                       & $L^\ast$   & 2.316 & 2.285 & 2.261 & 2.241 & 2.236 & 2.215 & 2.204 & 2.192 & 2.188 & 2.180 \\
                       & UCL$^\ast$ & 0.129 & 0.180 & 0.229 & 0.278 & 0.326 & 0.372 & 0.419 & 0.465 & 0.510 & 0.555 \\
                       & $L^{\ast\ast}$   & 1.584 & 1.794 & 1.874 & 1.936 & 1.945 & 1.960 & 1.968 & 1.973 & 1.971 & 1.965 \\
\multicolumn{1}{c}{}   & UCL$^{\ast\ast}$ & 0.076 & 0.135 & 0.191 & 0.246 & 0.299 & 0.351 & 0.403 & 0.455 & 0.505 & 0.555 \\
 \hline
\multicolumn{1}{c}{15} & $L$   & 2.343 & 2.270 & 2.260 & 2.243 & 2.230 & 2.210 & 2.201 & 2.192 & 2.180 & 2.179 \\
                       & UCL & 0.071 & 0.128 & 0.183 & 0.237 & 0.290 & 0.342 & 0.393 & 0.444 & 0.495 & 0.545 \\
                       & $L^\ast$   & 2.286 & 2.256 & 2.242 & 2.228 & 2.222 & 2.213 & 2.200 & 2.194 & 2.185 & 2.180 \\
                       & UCL$^\ast$ & 0.123 & 0.172 & 0.221 & 0.269 & 0.316 & 0.363 & 0.409 & 0.455 & 0.500 & 0.545 \\
                       & $L^{\ast\ast}$   & 1.562 & 1.772 & 1.871 & 1.921 & 1.942 & 1.954 & 1.962 & 1.965 & 1.971 & 1.963 \\
                       & UCL$^{\ast\ast}$ & 0.071 & 0.128 & 0.183 & 0.237 & 0.290 & 0.342 & 0.393 & 0.444 & 0.495 & 0.545 \\
 \hline
\multicolumn{1}{c}{20} & $L$   & 2.323 & 2.267 & 2.247 & 2.236 & 2.216 & 2.206 & 2.202 & 2.191 & 2.180 & 2.178 \\
                       & UCL & 0.068 & 0.124 & 0.179 & 0.232 & 0.284 & 0.336 & 0.388 & 0.438 & 0.489 & 0.539 \\
                       & $L^\ast$   & 2.284 & 2.250 & 2.235 & 2.220 & 2.215 & 2.205 & 2.197 & 2.186 & 2.185 & 2.178 \\
                       & UCL$^\ast$ & 0.119 & 0.168 & 0.216 & 0.263 & 0.310 & 0.357 & 0.403 & 0.448 & 0.494 & 0.539 \\
                       & $L^{\ast\ast}$   & 1.552 & 1.764 & 1.857 & 1.912 & 1.932 & 1.950 & 1.962 & 1.962 & 1.965 & 1.962 \\
\multicolumn{1}{c}{}   & UCL$^{\ast\ast}$ & 0.068 & 0.124 & 0.179 & 0.232 & 0.284 & 0.336 & 0.388 & 0.438 & 0.489 & 0.539\\
 \hline
\end{tabular}
\end{table}

\begin{table}
\caption{Simulation results ARL$_0=370$, $\lambda = 0.05$, and $\pi=0.99$.} \label{Sim_370_0.05_0.99}
\begin{tabular}{lccccccccccc}
 \hline
\multicolumn{1}{c}{}   & $p_0$   & 0.050  & 0.100   & 0.150  & 0.200   & 0.250  & 0.300   & 0.350  & 0.400   & 0.450  & 0.500   \\
                       & $p_0^\ast$  & 0.059 & 0.108 & 0.157 & 0.206 & 0.255 & 0.304 & 0.353 & 0.402 & 0.451 & 0.500   \\
\multicolumn{1}{c}{$n$}  & $p_0^{\ast\ast}$   & 0.050  & 0.100   & 0.150  & 0.200   & 0.250  & 0.300   & 0.350  & 0.400   & 0.450  & 0.500   \\
 \hline
\multicolumn{1}{c}{5}  & $L$   & 2.463 & 2.355 & 2.290 & 2.281 & 2.260 & 2.242 & 2.222 & 2.205 & 2.195 & 2.185 \\
                       & UCL & 0.088 & 0.151 & 0.209 & 0.265 & 0.320 & 0.374 & 0.426 & 0.477 & 0.528 & 0.578 \\
                       & $L^\ast$   & 2.432 & 2.341 & 2.306 & 2.276 & 2.257 & 2.237 & 2.216 & 2.203 & 2.195 & 2.184 \\
                       & UCL$^\ast$ & 0.100 & 0.160 & 0.217 & 0.272 & 0.325 & 0.378 & 0.429 & 0.479 & 0.529 & 0.578 \\
                       & $L^{\ast\ast}$   & 2.228 & 2.228 & 2.205 & 2.215 & 2.200 & 2.185 & 2.170 & 2.158 & 2.154 & 2.139 \\
                       & UCL$^{\ast\ast}$ & 0.088 & 0.151 & 0.209 & 0.265 & 0.320 & 0.373 & 0.426 & 0.477 & 0.528 & 0.578 \\
 \hline
\multicolumn{1}{c}{10} & $L$   & 2.374 & 2.295 & 2.270 & 2.258 & 2.230 & 2.221 & 2.211 & 2.202 & 2.185 & 2.182 \\
                       & UCL & 0.076 & 0.135 & 0.191 & 0.246 & 0.299 & 0.352 & 0.403 & 0.455 & 0.505 & 0.555 \\
                       & $L^\ast$   & 2.361 & 2.295 & 2.257 & 2.250 & 2.230 & 2.215 & 2.206 & 2.198 & 2.190 & 2.182 \\
                       & UCL$^\ast$ & 0.087 & 0.144 & 0.199 & 0.252 & 0.304 & 0.356 & 0.406 & 0.457 & 0.506 & 0.555 \\
                       & $L^{\ast\ast}$   & 2.144 & 2.183 & 2.182 & 2.190 & 2.169 & 2.167 & 2.157 & 2.153 & 2.144 & 2.137 \\
\multicolumn{1}{c}{}   & UCL$^{\ast\ast}$ & 0.076 & 0.135 & 0.191 & 0.246 & 0.299 & 0.352 & 0.403 & 0.455 & 0.505 & 0.555 \\
 \hline
\multicolumn{1}{c}{15} & $L$   & 2.343 & 2.270 & 2.260 & 2.243 & 2.230 & 2.210 & 2.201 & 2.192 & 2.180 & 2.179 \\
                       & UCL & 0.071 & 0.128 & 0.183 & 0.237 & 0.290 & 0.342 & 0.393 & 0.444 & 0.495 & 0.545 \\
                       & $L^\ast$   & 2.324 & 2.267 & 2.251 & 2.232 & 2.225 & 2.216 & 2.206 & 2.196 & 2.187 & 2.180 \\
                       & UCL$^\ast$ & 0.082 & 0.137 & 0.191 & 0.243 & 0.295 & 0.346 & 0.397 & 0.447 & 0.496 & 0.545 \\
                       & $L^{\ast\ast}$   & 2.115 & 2.156 & 2.172 & 2.172 & 2.166 & 2.158 & 2.151 & 2.149 & 2.144 & 2.137 \\
                       & UCL$^{\ast\ast}$ & 0.071 & 0.128 & 0.183 & 0.237 & 0.290 & 0.342 & 0.393 & 0.444 & 0.495 & 0.545 \\
 \hline
\multicolumn{1}{c}{20} & $L$   & 2.323 & 2.267 & 2.247 & 2.236 & 2.216 & 2.206 & 2.202 & 2.191 & 2.180 & 2.178 \\
                       & UCL & 0.068 & 0.124 & 0.179 & 0.232 & 0.284 & 0.336 & 0.388 & 0.438 & 0.489 & 0.539 \\
                       & $L^\ast$   & 2.314 & 2.267 & 2.247 & 2.228 & 2.218 & 2.211 & 2.204 & 2.193 & 2.185 & 2.178 \\
                       & UCL$^\ast$ & 0.079 & 0.133 & 0.186 & 0.238 & 0.290 & 0.340 & 0.391 & 0.440 & 0.490 & 0.539 \\
                       & $L^{\ast\ast}$   & 2.102 & 2.144 & 2.159 & 2.167 & 2.157 & 2.156 & 2.151 & 2.149 & 2.144 & 2.137 \\
\multicolumn{1}{c}{}   & UCL$^{\ast\ast}$ & 0.068 & 0.124 & 0.179 & 0.232 & 0.284 & 0.336 & 0.388 & 0.438 & 0.489 & 0.539\\
 \hline
\end{tabular}
\end{table}

\begin{table}
\caption{Simulation results ARL$_0=370$, $\lambda = 0.2$, and $\pi=0.95$.} \label{Sim_370_0.2_0.95}
\begin{tabular}{lccccccccccc}
 \hline
\multicolumn{1}{c}{}   & $p_0$   & 0.050  & 0.100   & 0.150  & 0.200   & 0.250  & 0.300   & 0.350  & 0.400   & 0.450  & 0.500   \\
                       & $p_0^\ast$  & 0.095 & 0.140  & 0.185 & 0.230  & 0.275 & 0.320  & 0.365 & 0.410  & 0.455 & 0.500   \\
\multicolumn{1}{c}{$n$}  & $p_0^{\ast\ast}$   & 0.050  & 0.100   & 0.150  & 0.200   & 0.250  & 0.300   & 0.350  & 0.400   & 0.450  & 0.500   \\
 \hline
\multicolumn{1}{c}{5}  & $L$   & 3.336 & 3.068 & 2.946 & 2.862 & 2.793 & 2.740 & 2.693 & 2.644 & 2.618 & 2.589 \\
                       & UCL & 0.158 & 0.237 & 0.307 & 0.371 & 0.430 & 0.487 & 0.541 & 0.593 & 0.644 & 0.693 \\
                       & $L^\ast$   & 3.078 & 2.976 & 2.883 & 2.807 & 2.769 & 2.722 & 2.686 & 2.629 & 2.603 & 2.588 \\
                       & UCL$^\ast$ & 0.230 & 0.294 & 0.352 & 0.406 & 0.459 & 0.509 & 0.558 & 0.603 & 0.648 & 0.693 \\
                       & $L^{\ast\ast}$   & 2.238 & 2.388 & 2.442 & 2.442 & 2.439 & 2.427 & 2.403 & 2.364 & 2.345 & 2.315 \\
                       & UCL$^{\ast\ast}$ & 0.159 & 0.237 & 0.307 & 0.370 & 0.430 & 0.488 & 0.542 & 0.593 & 0.643 & 0.692 \\
                        \hline
\multicolumn{1}{c}{10} & $L$   & 3.138 & 2.945 & 2.855 & 2.794 & 2.745 & 2.713 & 2.675 & 2.641 & 2.616 & 2.586 \\
                       & UCL & 0.122 & 0.193 & 0.257 & 0.318 & 0.375 & 0.431 & 0.484 & 0.536 & 0.587 & 0.636 \\
                       & $L^\ast$   & 2.960 & 2.878 & 2.793 & 2.767 & 2.718 & 2.696 & 2.660 & 2.642 & 2.609 & 2.587 \\
                       & UCL$^\ast$ & 0.186 & 0.245 & 0.299 & 0.353 & 0.403 & 0.453 & 0.500 & 0.547 & 0.592 & 0.636 \\
                       & $L^{\ast\ast}$   & 2.097 & 2.288 & 2.367 & 2.390 & 2.396 & 2.399 & 2.382 & 2.370 & 2.348 & 2.331 \\
\multicolumn{1}{c}{}   & UCL$^{\ast\ast}$ & 0.122 & 0.193 & 0.258 & 0.318 & 0.375 & 0.431 & 0.484 & 0.537 & 0.587 & 0.637 \\
 \hline
\multicolumn{1}{c}{15} & $L$   & 3.044 & 2.893 & 2.814 & 2.768 & 2.723 & 2.687 & 2.661 & 2.637 & 2.616 & 2.585 \\
                       & UCL & 0.107 & 0.175 & 0.236 & 0.295 & 0.351 & 0.406 & 0.459 & 0.511 & 0.562 & 0.611 \\
                       & $L^\ast$   & 2.897 & 2.832 & 2.775 & 2.735 & 2.705 & 2.686 & 2.656 & 2.635 & 2.617 & 2.587 \\
                       & UCL$^\ast$ & 0.168 & 0.225 & 0.278 & 0.329 & 0.379 & 0.428 & 0.475 & 0.522 & 0.567 & 0.611 \\
                       & $L^{\ast\ast}$   & 2.037 & 2.253 & 2.329 & 2.364 & 2.379 & 2.377 & 2.372 & 2.364 & 2.353 & 2.331 \\
                       & UCL$^{\ast\ast}$ & 0.107 & 0.175 & 0.236 & 0.295 & 0.352 & 0.406 & 0.459 & 0.511 & 0.562 & 0.611 \\
                        \hline
\multicolumn{1}{c}{20} & $L$   & 2.980 & 2.858 & 2.790 & 2.748 & 2.710 & 2.678 & 2.655 & 2.635 & 2.614 & 2.585 \\
                       & UCL & 0.098 & 0.164 & 0.224 & 0.282 & 0.337 & 0.391 & 0.444 & 0.496 & 0.547 & 0.596 \\
                       & $L^\ast$   & 2.859 & 2.796 & 2.756 & 2.726 & 2.691 & 2.674 & 2.654 & 2.632 & 2.612 & 2.586 \\
                       & UCL$^\ast$ & 0.157 & 0.212 & 0.265 & 0.316 & 0.365 & 0.413 & 0.460 & 0.506 & 0.552 & 0.596 \\
                       & $L^{\ast\ast}$   & 1.995 & 2.224 & 2.309 & 2.349 & 2.362 & 2.372 & 2.368 & 2.364 & 2.348 & 2.332 \\
\multicolumn{1}{c}{}   & UCL$^{\ast\ast}$ & 0.098 & 0.164 & 0.224 & 0.282 & 0.337 & 0.392 & 0.444 & 0.496 & 0.547 & 0.597\\
 \hline
\end{tabular}
\end{table}

\begin{table}
\caption{Simulation results ARL$_0=370$, $\lambda = 0.2$, and $\pi=0.99$.} \label{Sim_370_0.2_0.99}
\begin{tabular}{lccccccccccc}
 \hline
\multicolumn{1}{c}{}   & $p_0$   & 0.050  & 0.100   & 0.150  & 0.200   & 0.250  & 0.300   & 0.350  & 0.400   & 0.450  & 0.500   \\
                       & $p_0^\ast$  & 0.059 & 0.108 & 0.157 & 0.206 & 0.255 & 0.304 & 0.353 & 0.402 & 0.451 & 0.500   \\
\multicolumn{1}{c}{$n$}  & $p_0^{\ast\ast}$   & 0.050  & 0.100   & 0.150  & 0.200   & 0.250  & 0.300   & 0.350  & 0.400   & 0.450  & 0.500   \\
 \hline
\multicolumn{1}{c}{5}  & $L$   & 3.336 & 3.068 & 2.946 & 2.862 & 2.793 & 2.740 & 2.693 & 2.644 & 2.618 & 2.589 \\
                       & UCL & 0.158 & 0.237 & 0.307 & 0.371 & 0.430 & 0.487 & 0.541 & 0.593 & 0.644 & 0.693 \\
                       & $L^\ast$   & 3.257 & 3.061 & 2.935 & 2.842 & 2.792 & 2.748 & 2.693 & 2.633 & 2.603 & 2.588 \\
                       & UCL$^\ast$ & 0.173 & 0.250 & 0.316 & 0.377 & 0.436 & 0.492 & 0.545 & 0.594 & 0.644 & 0.693 \\
                       & $L^{\ast\ast}$   & 3.032 & 2.911 & 2.836 & 2.769 & 2.718 & 2.680 & 2.637 & 2.584 & 2.553 & 2.520 \\
                       & UCL$^{\ast\ast}$ & 0.159 & 0.237 & 0.307 & 0.370 & 0.430 & 0.488 & 0.542 & 0.593 & 0.643 & 0.692 \\
                        \hline
\multicolumn{1}{c}{10} & $L$   & 3.138 & 2.945 & 2.855 & 2.794 & 2.745 & 2.713 & 2.675 & 2.641 & 2.616 & 2.586 \\
                       & UCL & 0.122 & 0.193 & 0.257 & 0.318 & 0.375 & 0.431 & 0.484 & 0.536 & 0.587 & 0.636 \\
                       & $L^\ast$   & 3.095 & 2.913 & 2.845 & 2.789 & 2.741 & 2.712 & 2.670 & 2.648 & 2.612 & 2.588 \\
                       & UCL$^\ast$ & 0.136 & 0.203 & 0.266 & 0.325 & 0.381 & 0.435 & 0.488 & 0.539 & 0.588 & 0.636 \\
                       & $L^{\ast\ast}$   & 2.842 & 2.786 & 2.751 & 2.705 & 2.674 & 2.647 & 2.617 & 2.590 & 2.560 & 2.536 \\
\multicolumn{1}{c}{}   & UCL$^{\ast\ast}$ & 0.122 & 0.193 & 0.258 & 0.318 & 0.375 & 0.431 & 0.485 & 0.537 & 0.587 & 0.636 \\
 \hline
\multicolumn{1}{c}{15} & $L$   & 3.044 & 2.893 & 2.814 & 2.768 & 2.723 & 2.687 & 2.661 & 2.637 & 2.616 & 2.585 \\
                       & UCL & 0.107 & 0.175 & 0.236 & 0.295 & 0.351 & 0.406 & 0.459 & 0.511 & 0.562 & 0.611 \\
                       & $L^\ast$   & 3.001 & 2.873 & 2.806 & 2.761 & 2.720 & 2.687 & 2.657 & 2.639 & 2.619 & 2.587 \\
                       & UCL$^\ast$ & 0.120 & 0.185 & 0.245 & 0.302 & 0.357 & 0.410 & 0.462 & 0.513 & 0.563 & 0.611 \\
                       & $L^{\ast\ast}$   & 2.762 & 2.740 & 2.706 & 2.682 & 2.652 & 2.623 & 2.604 & 2.582 & 2.565 & 2.537 \\
                       & UCL$^{\ast\ast}$ & 0.107 & 0.175 & 0.236 & 0.295 & 0.352 & 0.406 & 0.459 & 0.511 & 0.562 & 0.611 \\
 \hline
\multicolumn{1}{c}{20} & $L$   & 2.980 & 2.858 & 2.790 & 2.748 & 2.710 & 2.678 & 2.655 & 2.635 & 2.614 & 2.585 \\
                       & UCL$_1$ & 0.098 & 0.164 & 0.224 & 0.282 & 0.337 & 0.391 & 0.444 & 0.496 & 0.547 & 0.596 \\
                       & $L^\ast$   & 2.948 & 2.841 & 2.781 & 2.736 & 2.707 & 2.677 & 2.655 & 2.634 & 2.613 & 2.584 \\
                       & UCL$^\ast$ & 0.111 & 0.174 & 0.232 & 0.288 & 0.343 & 0.396 & 0.448 & 0.498 & 0.548 & 0.596 \\
                       & $L^{\ast\ast}$   & 2.701 & 2.705 & 2.683 & 2.662 & 2.637 & 2.617 & 2.598 & 2.578 & 2.561 & 2.540 \\
\multicolumn{1}{c}{}   & UCL$^{\ast\ast}$ & 0.098 & 0.164 & 0.224 & 0.282 & 0.337 & 0.392 & 0.444 & 0.496 & 0.547 & 0.597\\
 \hline
\end{tabular}
\end{table}


\clearpage

\clearpage

  \renewcommand{\arraystretch}{0.90}
  \setlength\LTcapwidth{\linewidth}
\begin{longtable}{cccccccccccccccc}
\caption{Simulation results for ARL$_1$ based on Table~\ref{Sim_370_0.05_0.95}.} \label{ARL1_370_0.05_0.95}

\\
\hline

& & $p_1$    & 0.055  & 0.110  & 0.165  & 0.220  & 0.275  & 0.330  & 0.385  & 0.440  & 0.495  & 0.550  \\
$\delta$ & $n$ & $p_1^\ast$   & 0.100 & 0.149 & 0.199 & 0.248 & 0.298 & 0.347 & 0.397 & 0.446 & 0.496 & 0.545 \\
&& $p_1^{\ast\ast}$    & 0.055  & 0.110  & 0.165  & 0.220  & 0.275  & 0.330  & 0.385  & 0.440  & 0.495  & 0.550\\
 \hline
  0.1      &       5 & ARL$_1$ & 208.0  & 166.6 & 136.4  & 119.8 & 103.0  & 89.1  & 80.0   & 70.2  & 61.0   & 53.6  \\
                       &                     & ARL$_1^\ast$ & 251.9  & 198.4 & 160.6  & 137.1 & 119.7  & 103.8 & 90.9   & 79.3  & 69.9   & 61.9  \\
                       &                     & ARL$_1^{\ast\ast}$ & 207.7  & 166.0 & 138.8  & 120.8 & 103.8  & 88.0  & 80.6   & 70.4  & 61.8   & 54.0  \\
                       \cline{2-13}
                       & 10 & ARL$_1$ & 168.3  & 123.6 & 101.7  & 82.4  & 69.7   & 60.0  & 51.2   & 44.8  & 37.8   & 33.4  \\
                       &                     & ARL$_1^\ast$ & 218.6  & 154.0 & 121.2  & 98.7  & 81.7   & 68.9  & 60.2   & 51.0  & 43.8   & 38.8  \\
                       &                     & ARL$_1^{\ast\ast}$ & 168.4  & 125.5 & 100.5  & 82.9  & 69.9   & 59.9  & 51.1   & 44.4  & 38.1   & 33.4  \\
\cline{2-13}
                       & 15 & ARL$_1$ & 146.3  & 102.4 & 80.4   & 65.6  & 54.5   & 45.5  & 38.9   & 33.4  & 28.0   & 24.5  \\
                       &                     & ARL$_1^\ast$ & 196.2  & 130.9 & 99.8   & 79.7  & 63.5   & 53.7  & 45.6   & 39.2  & 33.6   & 28.8  \\
                       &                     & ARL$_1^{\ast\ast}$ & 144.6  & 103.6 & 80.9   & 65.7  & 54.0   & 45.7  & 38.9   & 33.5  & 28.5   & 24.6  \\
                       \cline{2-13}
                       & 20 & ARL$_1$ & 128.9  & 88.8  & 67.6   & 53.6  & 44.8   & 37.1  & 31.5   & 26.9  & 22.9   & 19.8  \\
                       &                     & ARL$_1^\ast$ & 176.1  & 115.9 & 84.1   & 65.2  & 52.8   & 43.6  & 37.7   & 31.9  & 26.8   & 23.1  \\
                       &                     & ARL$_1^{\ast\ast}$ & 128.5  & 88.8  & 67.1   & 53.3  & 44.8   & 37.1  & 31.5   & 26.9  & 22.9   & 19.8  \\
                        \hline
& & $p_1$    & 0.060   & 0.120  & 0.180   & 0.240  & 0.300    & 0.360  & 0.420   & 0.480  & 0.540   & 0.600   \\
                       &                     & $p_1^\ast$   & 0.104  & 0.158 & 0.212  & 0.266 & 0.320   & 0.374 & 0.428  & 0.482 & 0.536  & 0.590  \\
& & $p_1^{\ast\ast}$    & 0.060   & 0.120  & 0.180   & 0.240  & 0.300    & 0.360  & 0.420   & 0.480  & 0.540   & 0.600   \\
 \hline
0.2& 5& ARL$_1$ & 128.7  & 89.2  & 66.7   & 54.8  & 44.9   & 36.6  & 32.3   & 27.1  & 23.1   & 19.7  \\
                       &        & ARL$_1^\ast$ & 175.6  & 116.0 & 83.5   & 65.8  & 54.3   & 44.4  & 37.8   & 31.9  & 27.0   & 23.2  \\
                       &     & ARL$_1^{\ast\ast}$ & 127.2  & 89.0  & 68.0   & 55.0  & 45.0   & 36.3  & 32.2   & 27.3  & 23.3   & 19.7  \\
                       \cline{2-13}
                       & 10 & ARL$_1$ & 91.8   & 58.4  & 43.9   & 33.2  & 27.3   & 22.8  & 18.8   & 16.2  & 13.4   & 11.8  \\
                       &                     & ARL$_1^\ast$ & 136.2  & 78.3  & 55.1   & 42.2  & 32.5   & 26.9  & 22.7   & 18.7  & 15.7   & 13.8  \\
                       &                     & ARL$_1^{\ast\ast}$ & 90.9   & 59.1  & 43.7   & 33.2  & 27.2   & 22.9  & 18.8   & 16.2  & 13.5   & 11.8  \\
                       \cline{2-13}
                       & 15 & ARL$_1$ & 73.0   & 44.9  & 32.5   & 25.1  & 20.2   & 16.6  & 14.0   & 12.0  & 9.8    & 8.6   \\
                       &                     & ARL$_1^\ast$ & 114.1  & 62.5  & 42.3   & 31.8  & 24.1   & 19.9  & 16.5   & 14.1  & 12.0   & 10.0  \\
                       &                     & ARL$_1^{\ast\ast}$ & 72.4   & 45.3  & 32.3   & 25.1  & 20.1   & 16.7  & 14.0   & 12.0  & 9.9    & 8.6   \\
                       \cline{2-13}
                       & 20 & ARL$_1$ & 61.3   & 37.1  & 26.1   & 19.9  & 16.4   & 13.2  & 11.2   & 9.4   & 8.1    & 6.8   \\
                       &                     & ARL$_1^\ast$ & 96.9   & 51.9  & 34.2   & 25.2  & 19.6   & 15.8  & 13.5   & 11.3  & 9.3    & 8.0   \\
                       &                     & ARL$_1^{\ast\ast}$ & 60.9   & 37.0  & 26.1   & 19.8  & 16.3   & 13.3  & 11.2   & 9.4   & 8.1    & 6.8   \\
                        \hline

\end{longtable}

\clearpage

  \renewcommand{\arraystretch}{0.9}
  \setlength\LTcapwidth{\linewidth}
\begin{longtable}{cccccccccccccccc}
\caption{Simulation results for ARL$_1$ based on Table~\ref{Sim_370_0.05_0.99}.} \label{ARL1_370_0.05_0.99}
\\
\hline
& & $p_1$    & 0.055  & 0.110  & 0.165  & 0.220  & 0.275  & 0.330  & 0.385  & 0.440  & 0.495  & 0.550  \\
  $\delta$ & $n$   & $p_1^\ast$   & 0.064     & 0.118 & 0.172 & 0.226 & 0.280 & 0.334 & 0.387 & 0.441 & 0.495 & 0.549 \\
 & & $p_1^{\ast\ast}$    & 0.055  & 0.110  & 0.165  & 0.220  & 0.275  & 0.330  & 0.385  & 0.440  & 0.495  & 0.550  \\ 
\hline
 0.1   &  5   & ARL$_1$ & 209.4                         & 165.7  & 137.2  & 119.5  & 103.7  & 89.1   & 80.3   & 69.7   & 61.5   & 54.2  \\
                        &                      & ARL$_1^\ast$ & 217.0                         & 170.3  & 143.6  & 122.8  & 106.1  & 92.0   & 81.8   & 71.8   & 62.6   & 55.1  \\
                        &   & ARL$_1^{\ast\ast}$ & 206.5 & 165.2  & 136.6  & 120.2  & 104.1  & 88.8   & 79.9   & 69.7   & 61.6   & 53.8  \\
\cline{2-13}
                        &    10     & ARL$_1$ & 169.3                         & 124.5  & 101.4  & 82.8   & 69.9   & 60.3   & 51.0   & 44.8   & 37.9   & 33.3  \\
                        &                      & ARL$_1^\ast$ & 183.8                         & 130.4  & 103.9  & 83.9   & 72.2   & 61.5   & 52.5   & 45.6   & 39.1   & 34.4  \\
                        &   & ARL$_1^{\ast\ast}$ & 166.3 & 126.0  & 101.7  & 82.4   & 69.6   & 60.1   & 50.7   & 44.5   & 37.9   & 33.3  \\
\cline{2-13}
                        &   15    & ARL$_1$ & 146.2                         & 102.5  & 80.9   & 65.3   & 54.4   & 45.6   & 38.8   & 33.5   & 28.1   & 24.5  \\
                        &                      & ARL$_1^\ast$ & 158.1                         & 108.9  & 84.1   & 66.9   & 55.9   & 47.3   & 40.5   & 34.8   & 29.1   & 25.3  \\
                     &   & ARL$_1^{\ast\ast}$ & 145.6                         & 103.2  & 80.3   & 65.0   & 54.0   & 45.7   & 39.0   & 33.6   & 28.3   & 24.6  \\
\cline{2-13}
                        &   20  & ARL$_1$ & 130.0                         & 89.2   & 67.3   & 53.4   & 44.7   & 37.1   & 31.6   & 26.7   & 22.9   & 19.6  \\
                        &                      & ARL$_1^\ast$ & 140.6                         & 94.1   & 70.6   & 56.7   & 46.1   & 38.4   & 32.5   & 27.7   & 23.7   & 20.3  \\
 &   & ARL$_1^{\ast\ast}$ & 129.6                         & 87.9   & 67.4   & 53.3   & 44.8   & 36.8   & 31.6   & 26.9   & 23.0   & 19.7  \\
\hline
&& $p_1$    & 0.060   & 0.120  & 0.180   & 0.240  & 0.300    & 0.360  & 0.420   & 0.480  & 0.540   & 0.600   \\
                        &                      & $p_1^\ast$   & 0.069                        & 0.128 & 0.186 & 0.245 & 0.304  & 0.363 & 0.422 & 0.480 & 0.539 & 0.598 \\
&& $p_1^{\ast\ast}$    & 0.060   & 0.120  & 0.180   & 0.240  & 0.300    & 0.360  & 0.420   & 0.480  & 0.540   & 0.600   \\
\hline
        0.2  &    5      & ARL$_1$ & 128.2                         & 89.0   & 66.7   & 54.7   & 45.0   & 36.6   & 32.2   & 27.2   & 23.1   & 19.8  \\
                        &                      & ARL$_1^\ast$ & 138.3                         & 93.1   & 71.4   & 57.0   & 46.7   & 37.7   & 33.0   & 27.9   & 23.8   & 20.4  \\
                     &    & ARL$_1^{\ast\ast}$ & 128.0                         & 88.7   & 67.2   & 54.9   & 44.8   & 36.5   & 32.2   & 27.2   & 23.3   & 19.7  \\
\cline{2-13}
                        &     10     & ARL$_1$ & 92.0                          & 58.6   & 43.8   & 33.3   & 27.2   & 22.8   & 18.8   & 16.2   & 13.4   & 11.8  \\
                        &                      & ARL$_1^\ast$ & 103.3                         & 62.6   & 45.4   & 34.3   & 28.4   & 23.5   & 19.4   & 16.6   & 13.9   & 12.2  \\
                     &   & ARL$_1^{\ast\ast}$ & 90.8                          & 58.9   & 43.8   & 33.4   & 27.1   & 22.7   & 18.7   & 16.2   & 13.4   & 11.8  \\
\cline{2-13}
                        &  15     & ARL$_1$ & 72.8                          & 44.9   & 32.4   & 25.0   & 20.2   & 16.7   & 14.0   & 12.0   & 9.8    & 8.6   \\
                        &                      & ARL$_1^\ast$ & 81.8                          & 48.3   & 34.3   & 26.0   & 20.9   & 17.3   & 14.5   & 12.4   & 10.2   & 8.8   \\
                     &   & ARL$_1^{\ast\ast}$ & 72.3                          & 44.8   & 32.4   & 25.2   & 20.1   & 16.6   & 14.0   & 12.0   & 9.8    & 8.6   \\
\cline{2-13}
                        &    20     & ARL$_1$ & 61.1                          & 37.0   & 26.2   & 19.9   & 16.3   & 13.2   & 11.2   & 9.5    & 8.0    & 6.8   \\
                        &                      & ARL$_1^\ast$ & 68.9                          & 39.9   & 27.6   & 21.4   & 17.0   & 13.8   & 11.6   & 9.7    & 8.2    & 7.0   \\
 &   & ARL$_1^{\ast\ast}$ & 61.1                          & 37.0   & 26.1   & 19.9   & 16.4   & 13.3   & 11.2   & 9.5    & 8.0    & 6.8  \\
\hline

\end{longtable}

\clearpage

  \renewcommand{\arraystretch}{0.9}
  \setlength\LTcapwidth{\linewidth}
\begin{longtable}{cccccccccccccccc}
\caption{Simulation results for ARL$_1$ based on Table~\ref{Sim_370_0.2_0.95}.} \label{ARL1_370_0.2_0.95}

\\
\hline
& & $p_1$    & 0.055  & 0.110  & 0.165  & 0.220  & 0.275  & 0.330  & 0.385  & 0.440  & 0.495  & 0.550  \\
  $\delta$   &      $n$    & $p_1^\ast$  & {0.100}    & {0.149} & {0.199} & {0.248} & {0.298} & {0.347} & {0.397} & {0.446} & {0.496} & {0.545} \\
& & $p_1^{\ast\ast}$    & 0.055  & 0.110  & 0.165  & 0.220  & 0.275  & 0.330  & 0.385  & 0.440  & 0.495  & 0.550  \\
\hline
 0.1        &   5   & ARL$_1$ & 242.9                         & 209.1                     & 184.0                      & 166.6                     & 148.2                      & 132.7                     & 119.0                      & 108.3                     & 98.0                       & 88.5                      \\
                        &                      & ARL$_1^\ast$ & 282.1                         & 239.6                     & 207.2                      & 185.2                     & 164.6                      & 148.1                     & 135.0                      & 120.6                     & 107.8                      & 100.1                     \\
                        &   & ARL$_1^{\ast\ast}$ & 245.9                         & 209.6                     & 185.5                      & 164.2                     & 147.5                      & 133.0                     & 120.2                      & 106.9                     & 96.7                       & 85.5                      \\
\cline{2-13}
                        &   10    & ARL$_1$ & 213.2                         & 171.9                     & 142.9                      & 124.2                     & 107.0                      & 92.9                      & 81.6                       & 70.4                      & 62.0                       & 53.0                      \\
                        &                      & ARL$_1^\ast$ & 257.3                         & 203.4                     & 166.7                      & 142.2                     & 124.2                      & 107.9                     & 93.9                       & 82.6                      & 71.6                       & 61.8                      \\
                        &  & ARL$_1^{\ast\ast}$ & 213.3                         & 172.0                     & 144.6                      & 124.5                     & 107.3                      & 93.0                      & 81.1                       & 70.9                      & 61.6                       & 53.1                      \\
\cline{2-13}
                        &   15    & ARL$_1$ & 190.4                         & 147.4                     & 120.9                      & 101.6                     & 85.0                       & 73.0                      & 62.5                       & 53.3                      & 45.3                       & 38.6                      \\
                        &                      & ARL$_1^\ast$ & 232.8                         & 178.9                     & 142.8                      & 118.6                     & 99.3                       & 85.6                      & 72.7                       & 62.8                      & 53.8                       & 45.6                      \\
                        &  & ARL$_1^{\ast\ast}$ & 190.1                         & 147.9                     & 120.3                      & 100.3                     & 85.3                       & 72.8                      & 62.5                       & 53.3                      & 45.2                       & 38.6                      \\
\cline{2-13}
                        &    20  & ARL$_1$ & 175.0                         & 130.6                     & 104.2                      & 85.1                      & 71.5                       & 59.3                      & 50.3                       & 42.8                      & 36.2                       & 29.9                      \\
                        &                      & ARL$_1^\ast$ & 223.3                         & 161.8                     & 126.4                      & 102.5                     & 83.2                       & 70.4                      & 60.2                       & 50.3                      & 42.8                       & 35.7                      \\
  &  & ARL$_1^{\ast\ast}$ & 174.1                         & 130.7                     & 104.4                      & 85.1                      & 71.0                       & 59.7                      & 50.2                       & 42.8                      & 35.8                       & 30.4                      \\
  \hline
&& $p_1$    & 0.060   & 0.120  & 0.180   & 0.240  & 0.300    & 0.360  & 0.420   & 0.480  & 0.540   & 0.600   \\
                        &                      & $p_1^\ast$  & {0.104}     & {0.158} & {0.212}  & {0.266} & {0.32}   & {0.374} & {0.428}  & {0.482} & {0.536}  & {0.59}  \\
&& $p_1^{\ast\ast}$    & 0.060   & 0.120  & 0.180   & 0.240  & 0.300    & 0.360  & 0.420   & 0.480  & 0.540   & 0.600   \\
\hline
  0.2        &    5   & ARL$_1$ & 166.8                         & 128.4                     & 102.4                      & 85.4                      & 70.6                       & 59.6                      & 50.4                       & 43.1                      & 36.7                       & 31.1                      \\
                        &                      & ARL$_1^\ast$ & 217.7                         & 159.7                     & 124.5                      & 101.8                     & 84.2                       & 70.7                      & 60.3                       & 50.5                      & 42.7                       & 37.1                      \\
                        &   & ARL$_1^{\ast\ast}$ & 169.7 & 128.6                     & 104.3                      & 84.3                      & 70.2                       & 59.6                      & 50.4                       & 42.4                      & 36.3                       & 30.1                      \\
\cline{2-13}
                        &   10    & ARL$_1$ & 132.2                         & 91.4                      & 68.0                       & 53.7                      & 43.0                       & 34.8                      & 29.0                       & 23.9                      & 20.0                       & 16.6                      \\
                        &                      & ARL$_1^\ast$ & 182.1                         & 119.2                     & 86.6                       & 66.2                      & 52.9                       & 43.0                      & 34.9                       & 28.9                      & 23.9                       & 19.9                      \\
                        & & ARL$_1^{\ast\ast}$ & 131.1                         & 90.5                      & 68.2                       & 53.8                      & 42.9                       & 34.9                      & 28.9                       & 23.9                      & 19.9                       & 16.7                      \\
\cline{2-13}
                        &   15    & ARL$_1$ & 108.9                         & 70.9                      & 51.8                       & 39.5                      & 31.0                       & 25.1                      & 20.5                       & 16.9                      & 13.8                       & 11.5                      \\
                        &                      & ARL$_1^\ast$ & 156.0                         & 96.4                      & 66.7                       & 49.6                      & 38.1                       & 30.7                      & 24.8                       & 20.4                      & 16.7                       & 13.9                      \\
                        &  & ARL$_1^{\ast\ast}$ & 108.9                         & 70.7                      & 51.7                       & 39.2                      & 30.9                       & 25.1                      & 20.4                       & 16.8                      & 13.8                       & 11.5                      \\
\cline{2-13}
                        &    20   & ARL$_1$ & 93.4                          & 58.1                      & 41.3                       & 30.8                      & 24.4                       & 19.4                      & 15.7                       & 13.0                      & 10.7                       & 8.9                       \\
                        &                      & ARL$_1^\ast$ & 141.0                         & 81.4                      & 55.1                       & 40.0                      & 30.0                       & 23.8                      & 19.4                       & 15.6                      & 12.9                       & 10.7                      \\
  &  & ARL$_1^{\ast\ast}$ & 93.6                          & 58.1                      & 41.3                       & 30.9                      & 24.4                       & 19.4                      & 15.8                       & 13.1                      & 10.8                       & 9.0  \\
\hline

\end{longtable}

\clearpage

  \renewcommand{\arraystretch}{0.9}
  \setlength\LTcapwidth{\linewidth}
\begin{longtable}{cccccccccccccccc}
\caption{Simulation results for ARL$_1$ based on Table~\ref{Sim_370_0.2_0.99}.} \label{ARL1_370_0.2_0.99}

\\
\hline
& & $p_1$    & 0.055  & 0.110  & 0.165  & 0.220  & 0.275  & 0.330  & 0.385  & 0.440  & 0.495  & 0.550  \\
 $\delta$    &  $n$ & $p_1^\ast$  & 0.064     & 0.118 & 0.172 & 0.226 & 0.280 & 0.334 & 0.387 & 0.441 & 0.495 & 0.549 \\
& & $p_1^{\ast\ast}$    & 0.055  & 0.110  & 0.165  & 0.220  & 0.275  & 0.330  & 0.385  & 0.440  & 0.495  & 0.550  \\
 \hline
                        &   5  & ARL$_1$ & 242.0                         & 208.2                         & 184.5                      & 166.5                      & 147.2                      & 131.9                      & 118.8                      & 109.1                      & 98.7                       & 87.8                      \\
                        &                      & ARL$_1^\ast$ & 258.2                         & 217.8                         & 190.0                      & 167.5                      & 151.6                      & 138.4                      & 122.5                      & 109.2                      & 97.9                       & 90.4                      \\
                        &  & ARL$_1^{\ast\ast}$ & 244.2                         & 211.1                         & 185.5                      & 164.7                      & 147.0                      & 133.1                      & 120.0                      & 107.4                      & 96.0                       & 85.5                      \\
\cline{2-13}
                        &    10  & ARL$_1$ & 214.5                         & 172.2                         & 142.7                      & 124.0                      & 106.6                      & 93.4                       & 81.9                       & 70.7                       & 62.1                       & 53.0                      \\
                        &                      & ARL$_1^\ast$ & 222.0                         & 175.9                         & 147.9                      & 128.0                      & 110.7                      & 95.9                       & 83.6                       & 73.4                       & 63.4                       & 55.0                      \\
                        &  & ARL$_1^{\ast\ast}$ & 213.4 & 171.0 & 143.8                      & 123.2                      & 107.5                      & 93.0                       & 81.6                       & 71.1                       & 61.6                       & 53.3                      \\
\cline{2-13}
                        &   15   & ARL$_1$ & 190.0                         & 146.8                         & 120.6                      & 101.3                      & 85.4                       & 72.8                       & 62.1                       & 53.2                       & 45.2                       & 38.2                      \\
                        &                      & ARL$_1^\ast$ & 201.4                         & 154.7                         & 125.5                      & 104.5                      & 87.7                       & 75.2                       & 64.1                       & 55.3                       & 46.8                       & 39.7                      \\
                        &  & ARL$_1^{\ast\ast}$ & 191.0                         & 147.9                         & 120.4                      & 101.3                      & 85.2                       & 73.0                       & 62.7                       & 53.2                       & 45.4                       & 38.5                      \\
\cline{2-13}
                        &  20  & ARL$_1$ & 173.9                         & 131.1                         & 103.7                      & 85.6                       & 71.4                       & 59.4                       & 50.1                       & 42.6                       & 36.0                       & 30.1                      \\
                        &                      & ARL$_1^\ast$ & 185.6                         & 137.6                         & 108.2                      & 89.1                       & 74.4                       & 61.7                       & 52.5                       & 44.2                       & 37.3                       & 31.1                      \\
 & & ARL$_1^{\ast\ast}$ & 173.1                         & 130.3                         & 104.4                      & 85.1                       & 70.7                       & 59.6                       & 50.3                       & 42.7                       & 35.9                       & 30.2                      \\
 \hline
&& $p_1$    & 0.060   & 0.120  & 0.180   & 0.240  & 0.300    & 0.360  & 0.420   & 0.480  & 0.540   & 0.600   \\
                        &                      & $p_1^\ast$  & 0.069                        & 0.128 & 0.186 & 0.245 & 0.304  & 0.363 & 0.422 & 0.480 & 0.539 & 0.598 \\
&& $p_1^{\ast\ast}$    & 0.060   & 0.120  & 0.180   & 0.240  & 0.300    & 0.360  & 0.420   & 0.480  & 0.540   & 0.600   \\
 \hline
    0.2       &    5  & ARL$_1$ & 166.6                         & 129.0                         & 103.1                      & 85.7                       & 70.2                       & 59.4                       & 50.0                       & 43.0                       & 36.7                       & 31.0                      \\
                        &                      & ARL$_1^\ast$ & 183.6                         & 135.5                         & 108.4                      & 87.4                       & 72.8                       & 62.5                       & 52.2                       & 43.8                       & 37.3                       & 32.0                      \\
                        &   & ARL$_1^{\ast\ast}$ & 169.0                         & 129.1                         & 102.7                      & 84.8                       & 70.3                       & 59.6                       & 50.3                       & 42.5                       & 36.0                       & 30.2                      \\
\cline{2-13}
                        &    10  & ARL$_1$ & 132.4                         & 91.1                          & 67.6                       & 54.0                       & 43.0                       & 34.9                       & 29.0                       & 23.8                       & 20.0                       & 16.6                      \\
                        &                      & ARL$_1^\ast$ & 141.0                         & 95.2                          & 70.9                       & 55.7                       & 44.7                       & 36.1                       & 30.1                       & 25.0                       & 20.7                       & 17.2                      \\
                        &  & ARL$_1^{\ast\ast}$ & 129.9                         & 90.5                          & 68.0                       & 53.8                       & 43.2                       & 34.7                       & 29.0                       & 23.9                       & 20.0                       & 16.5                      \\
\cline{2-13}
                        &    15   & ARL$_1$ & 108.7                         & 70.6                          & 51.6                       & 39.5                       & 31.1                       & 24.9                       & 20.5                       & 16.9                       & 13.8                       & 11.5                      \\
                        &                      & ARL$_1^\ast$ & 119.9                         & 76.1                          & 54.2                       & 41.3                       & 32.5                       & 26.1                       & 21.1                       & 17.6                       & 14.4                       & 11.9                      \\
                        & & ARL$_1^{\ast\ast}$ & 109.5                         & 71.0                          & 51.7                       & 39.5                       & 31.0                       & 25.0                       & 20.5                       & 16.9                       & 13.8                       & 11.6                      \\
\cline{2-13}
                        &   20  & ARL$_1$ & 92.8                          & 58.1                          & 41.4                       & 31.0                       & 24.3                       & 19.3                       & 15.7                       & 13.0                       & 10.7                       & 8.9                       \\
                        &                      & ARL$_1^\ast$ & 104.3                         & 62.9                          & 43.6                       & 32.9                       & 25.6                       & 20.2                       & 16.4                       & 13.5                       & 11.2                       & 9.3                       \\
 &  & ARL$_1^{\ast\ast}$ & 92.8                          & 57.9                          & 41.3                       & 30.9                       & 24.5                       & 19.3                       & 15.8                       & 13.0                       & 10.8                       & 9.0  \\
\hline

\end{longtable}


      \begin{table}[!ht]
\caption{Numerical results based on naive estimator.}
\label{RDA}
\normalsize

 \centering
  \renewcommand{\arraystretch}{0.9} 
 \begin{tabular}{c ccccccccc}
 \hline
 & \multicolumn{2}{c}{Naive} & &\multicolumn{2}{c}{Correct ($\pi=0.95$)} && \multicolumn{2}{c}{Correct ($\pi=0.99$)} \\ \cline{2-3} \cline{5-6} \cline{8-9}
$\lambda$ & $L^\ast$ & UCL$^\ast$ && $L^{\ast\ast}$ & UCL$^{\ast\ast}$ && $L^{\ast\ast}$ & UCL$^{\ast\ast}$   \\
\hline 
0.05 & 2.222 & 0.126 &&  2.013 & 0.080 && 2.183 & 0.118 \\
     \hline
0.20 & 2.753 & 0.151 && 2.019 & 0.101 && 2.616 & 0.142 \\
     \hline
\end{tabular}

\end{table}

\clearpage


\clearpage 

\section*{References}

\refmark Argoti, M.A. and Carri\'on‐Garc\'ia, A. (2019). A heuristic method for obtaining quasi ARL‐unbiased p‐Charts. {\em Quality and Reliability Engineering International}, 35, 47–61.

\refmark Asif, F., Khan, S., and Noor-ul-Amin, M. (2020). Hybrid exponentially weighted moving average control chart with measurement error. {\em Iranian Journal of Science and Technology, Transaction A}, 44, 801-811.

\refmark Aslam, M., Rao, G. S., AL-Marshadi, A. H., and Jun, C.-H. (2019). A nonparametric HEWMA-$p$ control chart for variance in monitoring processes. {\em Symmetry}, 11, 356.

\refmark Bourke, P. D. (2006). The RL$_2$ chart versus the $np$ chart for detecting upward shifts in fraction defective. {\em Journal of Applied Statistics}, 33, 1-15.

\refmark Chen, L.-P. (2020).
Semiparametric estimation for the transformation model with length-biased data and covariate measurement error.  {\em Journal of Statistical Computation and Simulation}, 90, 420-442. 

\refmark Chen, L.-P. and Yi, G. Y. (2021). Analysis of noisy survival data with graphical proportional hazards measurement error models. {\em Biometrics}, 77, 956–969.

\refmark Chen, L.-P., Muzayyanah, S., Yang, S.-F., Wang, B., and Jiang, T.-A. (2022). Monitoring process location and dispersion simultaneously using a control region. {\em DYNA}, 97, 71-78.

\refmark Huwang, L. and Hung, Y. (2007). Effect of measurement error on monitoring multivariate process variability. {\em Statistica Sinica}, 17, 749-760.

\refmark Linna, K. W. and Woodall, W. H. (2001). Effect of measurement error on shewhart control charts. {\em Journal of Quality Technology}, 33, 213-222.

\refmark Linna, K. W. and Woodall, W. H., and Busby, K. L. (2001). The performance of multivariate control charts in the presence of measurement error. {\em Journal of Quality Technology}, 33, 349-355.

\refmark Maravelakis,  P., Panaretos, J., and  Psarakis, S. (2004). EWMA chart and measurement error. {\em Journal of Applied Statistics}, 31, 445-455.

\refmark Nguyen, H. D., Tran, K. P., Celano, G., Maravelakis, P. E., and Castagliola, P. (2020). On the effect of themeasurement error on Shewhart t and EWMA t control charts. {\em The International Journal of Advanced Manufacturing Technology}, 107, 4317–4332.

\refmark Pendrill, L. R. (2014). Using measurement uncertainty in decision-making and conformity assessment. {\em Metrologia}, 51, S206-S218.

\refmark Puydarrieux, S., Pou, J. M., Leblond, L., Fischer, N., Allard, A., Feinberg, M., and El Guennouni, D. (2019). Role of measurement uncertainty in conformity assessment. {\em 19th International Congress of Metrology}, 16003.

\refmark Qiu, P. (2014). {\em Introduction to Statistical Process Control}. CRC Press, New York.

\refmark Shongwe, S. C., Malela-Majika, J.-C., and Castagliola, P. (2020). On monitoring the process mean of autocorrelated observations with measurement errors using the $w$-of-$w$ runs-rules scheme. {\em 
Quality and Reliability Engineering International}, 36, 1144–1160.

\refmark Shu, M.-H. and Wu, H. C. (2010). Monitoring imprecise fraction of conforming items using p control charts. {\em Journal of Applied Statistics}, 37, 1283-1297.

\refmark Sparks, R. (2017). Linking EWMA p charts and the risk adjustment control charts. {\em Quality and Reliability Engineering International}, 33, 617–636.

\refmark Tran, K. D., Nguyen, H. D., Nguyen, T. H., and Tran, K. P. (2021). Design of a variable sampling interval exponentially weighted moving average median control chart in presence of measurement error. {\em Quality and Reliability Engineering International}, 37, 374-390.

\refmark Yang, S.-F. and Arnold, B. C. (2016a). A new approach for monitoring process variance. {\em Journal of Statistical Computation and Simulation}, 86, 2749-2765.

\refmark Yang, S.-F. and Arnold, B. C. (2016b). Monitoring process variance using an ARL-unbiased EWMA-p control chart. {\em Quality and Reliability Engineering International}, 32 1227-1235.

\refmark Yang, S.-F. (2016). An improved distribution-free EWMA mean chart. {\em Communications in Statistics - Simulation and Computation}, 45, 1410-1427.

\refmark Yang, S.-F. and Wu S. (2017). A double sampling scheme for process variability monitoring. {\em Quality and Reliability Engineering International}, 33, 2193-2204.

\refmark Yi, G. Y. (2017). {\em Statistical Analysis with Measurement Error and Misclassication: Strategy, Method and Application}. Springer.

\refmark Zou, C. and Tsung, F. (2010). Likelihood ratio-based distribution-free EWMA control charts. {\em Journal of Quality Technology}, 42, 174-196.

\end{document}